\documentclass[aps,twocolumn,nofootinbib,superscriptaddress]{revtex4-2}

\usepackage[english]{babel}
\usepackage[utf8]{inputenc}
\usepackage[normalem]{ulem}
\usepackage{amsmath,amssymb,amsfonts,physics,bm,graphicx,xcolor,mathtools,dsfont,hyperref}
\hypersetup{
	colorlinks=true,
	linkcolor={cyan!50!black},
	citecolor={red!50!black},
	urlcolor={red!40!black}
	}
\graphicspath{{fig/}}
\definecolor{myred}{rgb}{0.85,0,0}
\definecolor{mygray}{rgb}{0.87, 0.87, 0.87}

\let\oldbibitem\bibitem 
\renewcommand{\bibitem}{
    \renewcommand{\doi}[1]{\texttt{\href{https://doi.org/##1}{doi:##1}}} 
    \let\bibitem\oldbibitem 
    \oldbibitem 
}

\begin{document}

\title{Graph theory approach to exceptional points in wave scattering}

\author{Stefano Scali}
\email{s.scali@exeter.ac.uk}
\affiliation{Department of Physics and Astronomy, University of Exeter, Stocker Road, Exeter EX4 4QL, United Kingdom}
\author{Janet Anders}
\affiliation{Department of Physics and Astronomy, University of Exeter, Stocker Road, Exeter EX4 4QL, United Kingdom}
\affiliation{Institute of Physics and Astronomy, University of Potsdam, Karl-Liebknecht-Str.~24-25, 14476 Potsdam, Germany}
\author{Simon A. R. Horsley}
\affiliation{Department of Physics and Astronomy, University of Exeter, Stocker Road, Exeter EX4 4QL, United Kingdom}

\begin{abstract}
In this paper, we use graph theory to solve wave scattering problems in the discrete dipole approximation. As a key result of this work, in the presence of active scatterers, we present a systematic method to find arbitrary large--order zero eigenvalue exceptional points (EPs). This is achieved by solving a set of non--linear equations that we interpret, in a graph theory picture, as vanishing sums of scattering events. We then show how the total field of the system responds to parameter perturbations at the EP. Finally, we investigate the sensitivity of the power output to imaginary perturbation in the design frequency. This perturbation can be employed to trade sensitivity for a different dissipation balance of the system. The purpose of the results of this paper is manifold. On the one hand, we aim to shed light on the link between graph theory and wave scattering. On the other hand, the results of this paper find application in all those settings where zero eigenvalue EPs play a unique role like in coherent perfect absorption (CPA) structures.
\end{abstract}

\maketitle


Although wave scattering is an elementary process and straightforward to picture, its analysis continues to fuel developments in electromagnetic and acoustic material research. While a small object (particle) scatters as a point source with a strength proportional to the applied field, larger objects scatter the wave between their constituent parts. This multiple scattering process is an infinite chain of possible scattering events, interfering to give the total field. This complicated interaction breaks the simple relationship between the applied and scattered wave amplitudes. From this complex interaction, several fields of research emerge including metamaterials~\cite{Kadic_2019}, photonic crystals~\cite{Pitruzzello_2018}, propagation and imaging through disordered media~\cite{Gigan_2022}, and random lasing~\cite{Cao_2003}. 

The last decade has seen a large body of research into wave scattering in non--Hermitian materials, originating from Bender's proposed parity--time symmetric extension to quantum mechanics~\cite{Bender_1998}. Non--Hermitian materials differ from ordinary matter in that they are usually driven, containing regions where the wave can be amplified, in addition to regions of absorption. This absorption and re--emission of wave energy provides much more control over the wave field compared to passive structures, demonstrated in designs for invisible and reflectionless media~\cite{Horsley_2015,Makris_2015,Horsley_2016}, cloaking~\cite{Sounas_2015}, one--way propagation~\cite{Yin_2013}, coherent perfect absorption~\cite{Chong_2010,Baranov_2017}, and disordered media without scattering~\cite{King_2017}. Although initially an obstacle, controlled wave amplification has now been demonstrated from GHz~\cite{Hu_2017} to optical frequencies~\cite{Feng_2017}, as well as in acoustics~\cite{Shi_2016,Rivet_2018,Cho_2020}.

In this work, we investigate the problem of designing non--Hermitian arrays of particles with controllable exceptional point degeneracies. Exceptional points (EPs) are peculiar to non--Hermitian materials where two or more modes of the system have both eigenvalues and eigenvectors that coalesce. They have attracted considerable interest~\cite{Miri_2019} exhibiting an apparently increased sensitivity to system perturbations~\cite{Wiersig_2016,Hodaei_2017}, with the degenerate modes transforming into one another after cycling the system parameters~\cite{Uzdin_2011,Berry_2011}. To the best of our knowledge, while extensive work has been done on higher--order exceptional points~\cite{Hodaei_2017,Nada_2017,Sayyad_2022}, no consistent method to find $N^\mathrm{th}$--order EPs in wave scattering systems has been presented yet. In this work, we provide a recipe based on graph theory for implementing an exceptional point of arbitrary order in a system of scattering particles. The resulting system exhibits scattering properties with an extreme sensitivity to small changes in the particles' positions.

Our graph theory approach is based on the discrete dipole approximation (DDA)~\cite{Draine_1994,Yurkin_2007}. This is an established method for calculating the field scattered from any configuration of $N$ particles. Originally introduced by Purcell to calculate the scattering from astrophysical dust~\cite{Purcell_1973}, this method is now commonly applied to, e.g., metamaterial design~\cite{Landy_2014,Capers_2021} and wave propagation in disordered media~\cite{Baker_2021} thanks to its vast range of validity~\cite{Zubko_2010}. By treating the particles as point sources, with a strength proportional to the incident field, the scattering problem can be solved self consistently determining the field on each particle. This requires the inversion of an $N\times N$ matrix, which rapidly becomes analytically intractable as the number of particles (scatterers) increases. Here, we provide a graph theory representation of this matrix inversion. We use this to understand the requirements on the scatterer parameters for the system to exhibit an exceptional point of arbitrary order, finding a remarkably simple picture in terms of vanishing sums of graphs related to different scattering events.

The paper is organized as follows: in Sec.~\ref{sec:dda}, we review the discrete dipole approximation (DDA). In Sec.~\ref{sec:graphtheory}, we show how to interpret DDA by means of graph theory. In Sec.~\ref{sec:weakstrong}, we derive the single scattering events and define orders of interactions. By means of the graph theory interpretation, we perform and give insights on weak and strong interaction limits. In Sec.~\ref{sec:nthorderEP}, we present a method to design $N^\mathrm{th}$--order EPs with zero eigenvalue in systems described by DDA, perhaps the most important result of this paper. To do this, we derive the conditions to find these EPs (Sec.~\ref{ssec:EPconditions}) and, consequently, we interpret these conditions in terms of graphs in a scattering setting (Sec.~\ref{ssec:graphEP}). In this setting, we show the effects of the EPs on the system's properties (Sec.~\ref{ssec:trading}), namely the total field and the power output. Finally, we show how one can exploit perturbations to the design resonant frequency to tune the dissipation balance across the array of scatterers. However, this comes at the cost of a broader power output. In Sec.~\ref{sec:conclusion}, we conclude by summarizing the results and possible next developments.


\section{Discrete dipole approximation}
\label{sec:dda}
\begin{figure}
    \centering
    \includegraphics[width=0.72\linewidth]{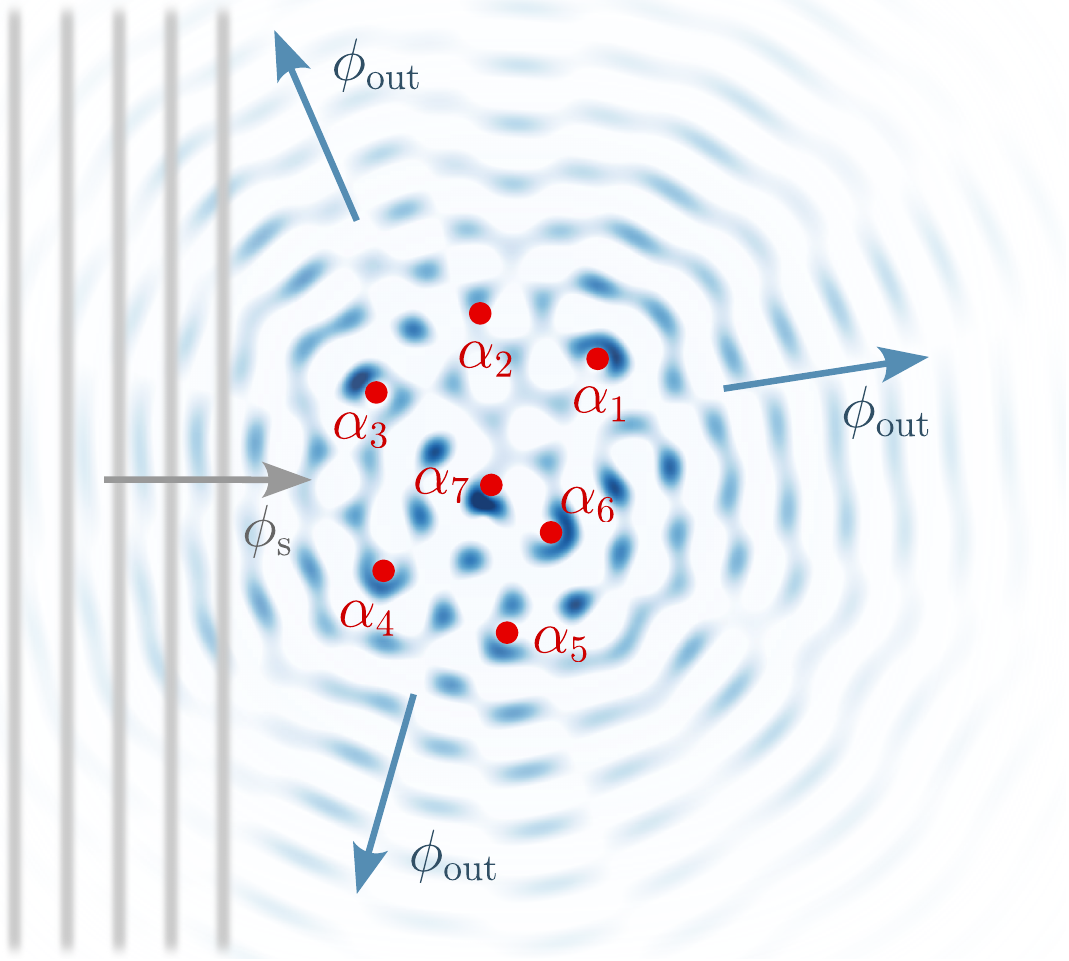}
    \caption{Schematic of a source field $\phi_\mathrm{s}$ incident onto an array of sub--wavelength size scatterers (red dots) with polarizabilities $\alpha_n$. The scatterers respond to the source field, producing an outgoing field $\phi_{\mathrm{out}}=\sum_n\phi_n^\mathrm{out}$.}
    \label{fig:system_dda}
\end{figure}
For simplicity, we restrict our theory to scalar waves of amplitude $\phi$ (e.g., the pressure of an acoustic wave in a fluid or, in two dimensions, the fundamental mode of a waveguide), although there is no obstacle to adapting our theory to vector waves. A model of the system presented in the following is shown in Fig.~\ref{fig:system_dda}. We take $N$ scattering particles of polarizability $\alpha_n$, with $n=1,2,\cdots,N$. Subject to an incoming wave of amplitude $\phi_{\mathrm{inc}}$, each of these particles will act as a point source $\mathrm{s}_{n}$ of strength
\begin{equation}\label{eq:source_n}
    \mathrm{s}_n(\bm{x})=\alpha_{n}\,\phi_{\mathrm{inc}}(\bm{x}_{n})\,\delta^{(3)}(\bm{x}-\bm{x}_n).
\end{equation}
Note that the incoming field $\phi_{\mathrm{inc}}(\bm{x}_{n})$ is defined as the total field at position $\bm{x}_n$ (position of the scatterer $\alpha_n$) minus the self--field of the scatterer.  The total field $\phi(\bm{x})$ obeys the three dimensional Helmholtz equation, including the sources of scattered waves given in Eq.~\eqref{eq:source_n},
\begin{equation}\label{eq:helmholtz}
    (\nabla^2+k_0^2)\phi(\bm{x}) = \sum_{n=1}^N\alpha_n\,\phi_\mathrm{inc}(\bm{x}_n)\,\delta^{(3)}(\bm{x}-\bm{x}_n)+\mathrm{s}(\bm{x}),
\end{equation}
where $k_0=\omega_0/c$ is the wavenumber with $\omega_0$ the resonant frequency, and $\mathrm{s}(\bm{x})$ is the externally driven source of waves in the system. Throughout this paper, we assume $c=1$. The solution to the Helmholtz equation \eqref{eq:helmholtz} can be written in terms of the 3D Green's function $\displaystyle G(\bm{x},\bm{x}_n)=-\exp(ik_0|\bm{x}-\bm{x}_n|)/(4\pi|\bm{x}-\bm{x}_n|)$, which is the solution to $(\nabla^2+k_0^2)G(\bm{x},\bm{x}_n) = \delta^{(3)}(\bm{x}-\bm{x}_n)$.  Integrating the Green function against the right hand side of Eq.~\eqref{eq:helmholtz} we have the solution to Eq.~\eqref{eq:helmholtz}, which takes the form
\begin{equation}\label{eq:phi_solution}
    \phi(\bm{x}) = \sum_{n=1}^N\alpha_n\,G(\bm{x},\bm{x}_n)\,\phi_\mathrm{inc}(\bm{x}_n) + \phi_\mathrm{s}(\bm{x}),
\end{equation}
where $\phi_\mathrm{s}(\bm{x})$ is the integral of the Green's function over the source $\mathrm{s}(\bm{x})$. To determine the unknowns $\phi_{\mathrm{inc}}(\bm{x}_n)$, Eq.~\eqref{eq:phi_solution} is evaluated on each of the $N$ scatterers, excluding the infinite self--field, and demanding self--consistency,
\begin{equation}\label{eq:self_consistency}
    \phi_\mathrm{inc}(\bm{x}_m) = \sum_{\substack{n=1\\n\neq m}}^N\alpha_n\,G(\bm{x}_m,\bm{x}_n)\,\phi_\mathrm{inc}(\bm{x}_n) + \phi_\mathrm{s}(\bm{x}_m).
\end{equation}
To write the problem in a more convenient form, we scale our field amplitudes by the polarizability, defining the new set of unknowns $\vspace{10pt}\Tilde{\phi}_{\mathrm{inc}}(\bm{x}_n) = \alpha_n\,\phi_{\mathrm{inc}}(\bm{x}_n)$. Writing Eqs.~\eqref{eq:self_consistency} in matrix form, the solution is
\begin{equation}\label{eq:matrix_form}
    \bm{M}^{-1}\bm{\phi}_\mathrm{s} = \Tilde{\bm{\phi}}_\mathrm{inc},
\end{equation}
where the \emph{interaction matrix} $\bm{M}$ is given by
\begin{align}\label{eq:M}
    \bm{M}=
    \begin{pmatrix}
        \alpha_{1}^{-1} & -G(\bm{x}_1,\bm{x}_2) & -G(\bm{x}_1,\bm{x}_3) & \dots \\[3pt]
        -G(\bm{x}_2,\bm{x}_1) & \alpha_{2}^{-1} & -G(\bm{x}_2,\bm{x}_3) & \dots \\[3pt]
        -G(\bm{x}_3,\bm{x}_1) & -G(\bm{x}_3,\bm{x}_2) & \alpha_{3}^{-1} & \dots \\[3pt]
        \vdots & \vdots & \vdots & \ddots
    \end{pmatrix},
\end{align}
with the source field vector $\bm{\phi}_\mathrm{s} = (\phi_\mathrm{s}(\bm{x}_1), \phi_\mathrm{s}(\bm{x}_2), \cdots, \phi_\mathrm{s}(\bm{x}_N))^T$, and the incident field vector $\Tilde{\bm{\phi}}_\mathrm{inc} = (\Tilde{\phi}_\mathrm{inc}(\bm{x}_1), \Tilde{\phi}_\mathrm{inc}(\bm{x}_2), \cdots, \Tilde{\phi}_\mathrm{inc}(\bm{x}_N))^T$.
Note that, in general, the matrix $\bm{M}$ is non--Hermitian, being both complex and symmetric. In non-reciprocal systems~\cite{Salary_2019}, the interaction matrix is both complex and asymmetric. From Eq.~\eqref{eq:matrix_form}, we can therefore find a solution for the incident fields $\Tilde{\bm{\phi}}_\mathrm{inc}$ and consequently the total field $\phi(\bm{x})$ using Eq.~\eqref{eq:phi_solution}. This is the discrete dipole approximation (DDA) method for solving scattering problems~\cite{DeVoe_1964,Purcell_1973,Draine_1994}, reducing the entire problem to the matrix inversion $\bm{M}^{-1}$. This must be done numerically even for a small number of scatterers~\cite{Yurkin_2007}.


\section{Graph theory interpretation of wave scattering}
\label{sec:graphtheory}
Graph theory is a branch of mathematics rooted in Euler's solution to the problem of the seven bridges of K\"onigsberg~\cite{Euler_1735}. From here, graph theory stemmed and evolved, finding applications to many problems in science and engineering~\cite{Foulds_1992}.

\begin{figure*}
    \centering
    \includegraphics[width=0.89\linewidth]{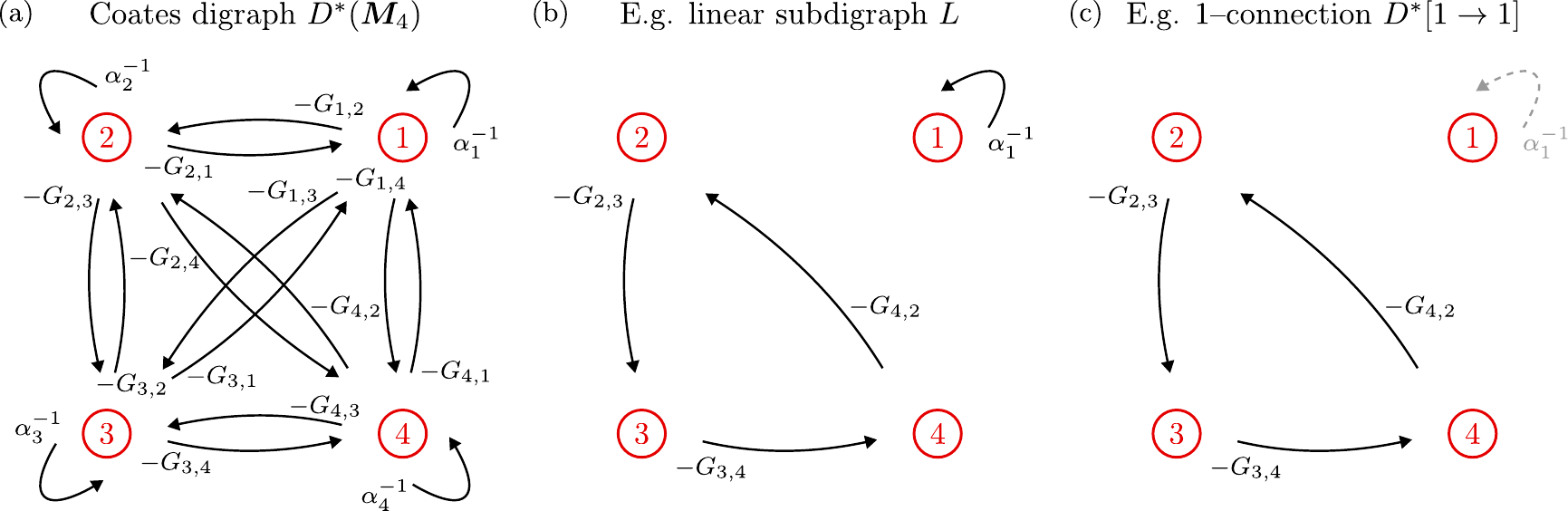}
    \caption{Example graphs used to describe the scattering system and the single scattering events. In panel (a), we show the Coates digraph $D^*(\bm{M}_4)$, representation of the matrix $\bm{M}_4$. The vertices represent the scatterers with the self--loops weighted by the inverse polarizabilities $\alpha_i^{-1}$ while the edges represent the interactions weighted by the Green's functions $G_{i,j}=G(\bm{x}_i,\bm{x}_j)$. Note that the labels of self--loops and edges are always placed as close as possible to the origin of the arrows they refer to. In panel (b), we show an example of linear subdigraph $L$ of $D^*(\bm{M}_4)$, i.e., a subdigraph in which exactly one edge enters and exactly one edge leaves each vertex. Summing the weights of all the linear subdigraphs of $D^*(\bm{M}_4)$, one obtains $\det(\bm{M}_4)$. In panel (c), we show an example of 1--connection $D^*[1\rightarrow 1]$ built from the linear subdigraph $L$. This is built by removing the edge $1\rightarrow 1$, as described in the main text. Summing the weights of all the 1--connections from $i$ to $j$, one obtains $\mathrm{adj}(\bm{M}_4)_{i,j}$.}
    \label{fig:k4}
\end{figure*}

The interaction matrix $\bm{M}$ in Eq.~\eqref{eq:M} can be represented as a graph (e.g., in panel (a) of Fig.~\ref{fig:k4}), where the diagonal elements (the particles' self-interaction $1/\alpha_i$) are represented as vertices, and their interaction ($-G(\bm{x}_i,\bm{x}_j)$) as edges. Multiple scattering events between the particles can thus be represented as a path on this graph, known as a Coates digraph. This representation links interactions and objects to edges and vertices respectively, fundamental constituents of any graph.

For example, take a 4--scatterer system whose matrix $\bm{M}_4$ is the $4\times 4$ equivalent of Eq.~\eqref{eq:M}. In panel (a) of Fig.~\ref{fig:k4}, we represent the matrix $\bm{M}_4$ as the complete Coates digraph $D^*(\bm{M}_4)$. Following convention~\cite{Coates_1959,Brualdi_2008}, we refer to the Coates digraph using a star superscript. The Coates digraph is constructed as follows: the scatterers are represented by vertices, the Green's function interactions take the role of the edges, and the intrinsic (inverse) polarizabilities of the single scatterers are identified by the vertices' self--loops. This graph earns the technical name of \emph{vertex--labeled directed weighted simple graph permitting loops}~\cite{West_2001,Bollob_s_1979}. From now on, we will shorten and refer to this type of graphs as digraphs or simply graphs.

This interpretation of the interaction matrix allows us to calculate the inversion of the matrix in Eq.~\eqref{eq:matrix_form} using graph theory. To do this, we consider the usual formula for the inversion of a matrix~\cite{Greub_1963},
\begin{equation}\label{eq:inversion}
    \bm{M}^{-1} = \frac{\mathrm{adj}(\bm{M})}{\det(\bm{M})} ,
\end{equation}
where $\mathrm{adj}(\bm{M})$ and $\det(\bm{M})$ are the adjugate (transpose of the cofactor matrix) and the determinant of $\bm{M}$, respectively. The $i,j$--th element of the adjugate matrix is defined as $\mathrm{adj}(\bm{M})_{i,j}=(-1)^{i+j}\det(\bm{M}_{(j,i)})$, where $\bm{M}_{(j,i)}$ is the minor\footnote{In this paper, we call a ``minor'' an $n\times n$ matrix built by removing $m$ rows and columns from an $N\times N$ matrix, with $N=m+n$. We will refer to the determinant of such a matrix as the ``determinant of a minor''.} built by removing the $j^\mathrm{th}$ row and the $i^\mathrm{th}$ column from the matrix $\bm{M}$. Therefore, both terms on the right hand side of Eq.~\eqref{eq:inversion} depend on determinant evaluations.

This form of inversion has a distinct interpretation in graph theory. It is thanks to this graph interpretation that we will be able to distinguish and identify different scattering events, ultimately solving for the total field of the system. In addition, using the same interpretation, we will illustrate a new visual way to build the condition to find zero eigenvalue EPs in scattering systems.

The determinant of a generic matrix $\bm{A}$ can be calculated using the Coates' determinant formula~\cite{Coates_1959,Greenman_1976,Brualdi_2008},
\begin{equation}\label{eq:det}
    \det(\bm{A}) = (-1)^N \sum_{L\in\mathcal{L}(\bm{A})}(-1)^{c(L)}\gamma(L) ,
\end{equation}
where $N$ is the number of vertices of the Coates digraph $D^*(\bm{A})$ and $L$ is an element in the set $\mathcal{L}(\bm{A})$ of all the possible linear subdigraphs of the Coates digraph $D^*(\bm{A})$~\cite{Coates_1959}. A linear subdigraph of the Coates digraph $D^*(\bm{A})$ is a subdigraph of $D^*(\bm{A})$ in which exactly one edge enters and exactly one edge leaves each vertex~\cite{Greenman_1976,Brualdi_2008}. The term $\gamma(L)$ is the product of the weights of the edges of $L$, and $c(L)$ is the number of cycles contained in $L$, i.e., the number of closed loops of the specific graph.

In panel (b) of Fig.~\eqref{fig:k4}, we show an example of a linear subdigraph $L$ of the Coates digraph $D^*(\bm{M}_4)$ (with $N=4$). Following the just mentioned definition, note that exactly one edge enters and leaves each vertex. The number of cycles of this graph is $c(L)=2$, while its weight is $\gamma(L)=-\alpha_1^{-1}G_{2,3}G_{3,4}G_{4,2}$. Following the same procedure applied in this example, we obtain the determinant of the matrix $\bm{A}$ by simply adding, according to Eq.~\eqref{eq:det}, the appropriately--signed weights of the linear subdigraphs of $D^*(\bm{A})$.

Using a similar construction, the expression for the adjugate of a generic matrix $\bm{A}$ is~\cite{Brualdi_2008},
\begin{equation}\label{eq:adj}
    \mathrm{adj}(\bm{A})_{i,j} = (-1)^N \sum_{D^*[i\rightarrow j]}(-1)^{c(D^*[i\rightarrow j])+1}\gamma(D^*[i\rightarrow j]) ,
\end{equation}
where the sum runs over all the possible 1--connections $D^*[i\rightarrow j]$ of the Coates digraph. A 1--connection $D^*[i\rightarrow j]$ is obtained from a linear subdigraph (containing the edge $j\rightarrow i$) by simply removing the edge $j\rightarrow i$. Note that, in the case $i=j$, this corresponds to removing the self--loop at vertex $i$.

An example 1--connection is shown in panel (c) of Fig.~\ref{fig:k4}. Starting by considering the linear subdigraph $L$ in panel (b), we remove the edge $j\rightarrow i$, that is, the self--loop $1\rightarrow 1$. In this way, we obtain the corresponding 1--connection having number of cycles $c(D^*[1\rightarrow 1])=1$ and weight $\gamma(D^*[1\rightarrow 1])=-G_{2,3}G_{3,4}G_{4,2}$. Following the same procedure applied in this example, we obtain the adjugate element $i,j$ of the matrix $\bm{A}$ by simply adding, according to Eq.~\eqref{eq:adj}, the appropriately--signed weights of the 1--connections of $D^*([i\rightarrow j])$. See appendix \ref{app:graphintro} for further examples and more formal definitions of Coates digraphs, linear subdigraphs, and 1--connections.

As a result, we can graphically represent Eqs.~\eqref{eq:adj} and \eqref{eq:det} for the matrix inversion \eqref{eq:inversion}, key for the evaluation of the total field of the system \eqref{eq:phi_solution}. These graph theory constructions, namely 1--connections and linear subdigraphs, give us a visual and systematic way of computing the elements of the inverse matrix $\bm{M}^{-1}$. I.e., each element $\displaystyle (\bm{M}^{-1})_{i,j}=\mathrm{adj}(\bm{M})_{i,j}/\det(\bm{M})$ is evaluated by dividing the weighted sum of the 1--connections from vertex $i$ to $j$ by the weighted sum of the linear subdigraphs of $D^*(\bm{M})$. As seen in section \ref{sec:dda}, this inverse allows us to solve for the total field of the system \eqref{eq:phi_solution}. Although graph theory doesn't reduce the number of calculations required to perform this inversion, it provides a intuitive representation of any scattering process in terms of a sequence of multiple scattering events. As we shall see, this allows us to give a graphical recipe for finding exceptional points in resonant scatterer arrays.


\section{Identification of different scattering orders}
\label{sec:weakstrong}
\begin{figure*}
    \centering
    \includegraphics[width=0.98\linewidth]{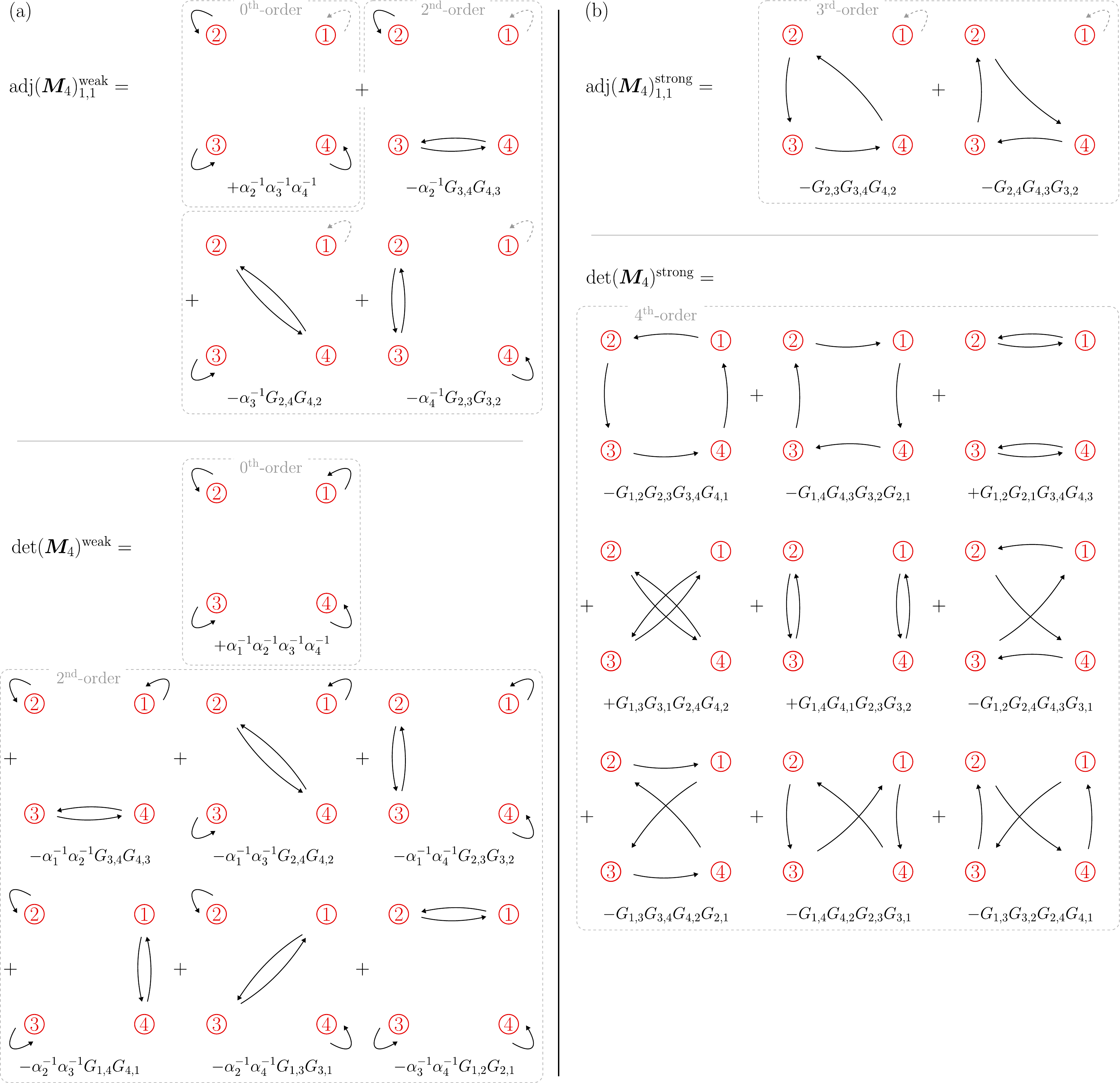}
    \caption{We show an example of construction of the element $(\bm{M}_4^{-1})_{1,1}=\mathrm{adj}(\bm{M}_4)_{1,1}/\det(\bm{M}_4)$ in the weak (panel (a)) and strong (panel (b)) approximations. While the weak approximation accounts for scattering events up to $2^\mathrm{nd}$--order in the interaction ($\propto G^2$), the strong approximation accounts for interactions of $4^\mathrm{th}$ ($\propto G^4$) or the highest non--trivial order. The contributions of the single graphs are derived using Eq.~\eqref{eq:adj} for the adjugate and Eq.~\eqref{eq:det} for the determinant. In the top panels, we show the 1--connections $D^*[1\rightarrow 1]$ obtained by removing the self--loop in vertex 1 from the linear subdigraphs $L$ that include the edge $1\rightarrow 1$. In the bottom panels, we show the linear subdigraphs obtained from the Coates digraph $D^*(\bm{M}_4)$.}
    \label{fig:graphs_approx}
\end{figure*}
Before treating the problem of exceptional points in these scatterer arrays, we show how we can use Eqs.~\eqref{eq:det} and \eqref{eq:adj} for the construction of the elements of the inverse matrix $\bm{M}^{-1}$ in the case of weak and strong interaction limits of the system. These limits are taken by controlling the order of magnitude of the distance between the scatterers relative to the magnitude of the wavenumber used to probe the system. This results in a change of the interaction terms in the form of Green's functions $G$. To show how to evaluate these limits, we firstly demonstrate how 1--connections and linear subdigraphs capture all the possible interaction paths of the signal in the system. This allow us to identify scattering events of different orders to build approximations.

As a simple example, we consider a system of two scatterers characterized by polarizabilities $\alpha_1$ and $\alpha_2$, symmetrically interacting via the Green's function $G_{1,2}$. Now, we constructively build all the possible paths (or scattering events) of the system. To do this, we evaluate the incident field on the first scatterer, $\phi_\mathrm{inc}(\bm{x}_1)$, while analogous considerations can be done for the second scatterer. The field $\phi_\mathrm{inc}(\bm{x}_1)$ is the sum of all the possible paths starting from the different scatterers of the system and ending in scatterer $1$. All these signals are scaled by the polarizability of the scatterer itself, $\alpha_1$. We start adding the contribution of a signal generated in scatterer $1$, $\phi_\mathrm{inc}(\bm{x}_1) = \left[\phi_\mathrm{s}(\bm{x}_1)\alpha_1 + \cdots\right]$, where the first term on the RHS is given by the source field. Proceeding in the same way, a signal propagating from the second scatterer is scaled by the polarizability of the scatterer itself, $\alpha_2$, then weighted by the interaction $G_{1,2}$ connecting the two scatterers, obtaining $\phi_\mathrm{inc}(\bm{x}_1) = \left[\phi_\mathrm{s}(\bm{x}_1)\alpha_1 + \phi_\mathrm{s}(\bm{x}_2)\alpha_2G_{1,2}\alpha_1\right]$. While these contributions account for the ``one--round trips'', the signals can propagate back and forth in the systems. Considering ``multiple--round trips'', we obtain
\begin{align}
    \phi_\mathrm{inc}(\bm{x}_1) = &\left[\phi_\mathrm{s}(\bm{x}_1)\alpha_1 + \phi_\mathrm{s}(\bm{x}_2)\alpha_2G_{1,2}\alpha_1 \right] \\ \nonumber
    &\left[1+\alpha_1\alpha_2G_{1,2}^2 + (\alpha_1\alpha_2G_{1,2}^2)^2 +\cdots\right] ,
\end{align}
where the term in the second square bracket accounts for the paths of different orders and extend to an infinite number of interactions. In the case of $|\alpha_1\alpha_2G_{1,2}^2|<1$, this last term can be written using the closed form of the geometric series as
\begin{equation}\label{eq:scattering}
    \phi_\mathrm{inc}(\bm{x}_1) = \frac{\left[\phi_\mathrm{s}(\bm{x}_1)\alpha_1 + \phi_\mathrm{s}(\bm{x}_2)\alpha_2G_{1,2}\alpha_1 \right]}{1 - \alpha_1\alpha_2G_{1,2}^2} .
\end{equation}
This is the analytical solution to Eq.~\eqref{eq:matrix_form} for the incident field $\phi_\mathrm{inc}(\bm{x}_1)$ in the case of a symmetric 2--scatterer system. Note that, in Eq.~\eqref{eq:scattering}, the terms in the numerator (i.e., the adjugate terms or 1--connections) represent the single scattering events, while the denominator (i.e., the determinant or linear subdigraphs) represent the possible multiple repetitions of the single scattering events. We identify the single scatter events and multiple repetitions by their order in the interaction $G$. For example, in Eq.~\eqref{eq:scattering}, the numerator is made of $0^\mathrm{th}$ and $1^\mathrm{st}$--order scattering events. In the same way, the denominator is made of $0^\mathrm{th}$--and $2^\mathrm{nd}$--order multiple repetitions. Proceeding in the same way for an arbitrary number of scatterers, we can build single scattering events and identify paths containing $i^\mathrm{th}$--order interactions.

Now, we translate this interpretation of scattering events into the graph theory picture of Sec.~\ref{sec:graphtheory} and we define different regimes of approximation. To do this, as a second example, we consider again the system described by $\bm{M}_4$. In Fig.~\ref{fig:graphs_approx}, we show how to evaluate the term $(\bm{M}_4)_{1,1}^{-1} = \mathrm{adj}(\bm{M}_4)_{1,1}/\det(\bm{M}_4)$ for the weak (panel (a)) and strong (panel (b)) coupling limits. By means of the construction shown above, in the case of weakly interacting scatterers, we restrict the sums in Eqs.~\eqref{eq:adj} and \eqref{eq:det} to those 1--connections/linear subdigraphs carrying weights $\gamma$ up to second order in the interactions $G$ (i.e., up to $G^2$), similar to the truncation of the Born series to second order~\cite{Newton_1982}. With this approximation, we account for all those scattering processes whose graphs include no more than 2 edges (self--loops excluded), as shown in Fig.~\ref{fig:graphs_approx} panel (a). Approximating both the adjugate terms and the full determinant of the matrix $\bm{M}_4$, we can evaluate the entries of $\bm{M}_4^{-1}$, as per Eq.~\eqref{eq:inversion}.
\begin{figure}
    \centering
    \includegraphics[width=\linewidth]{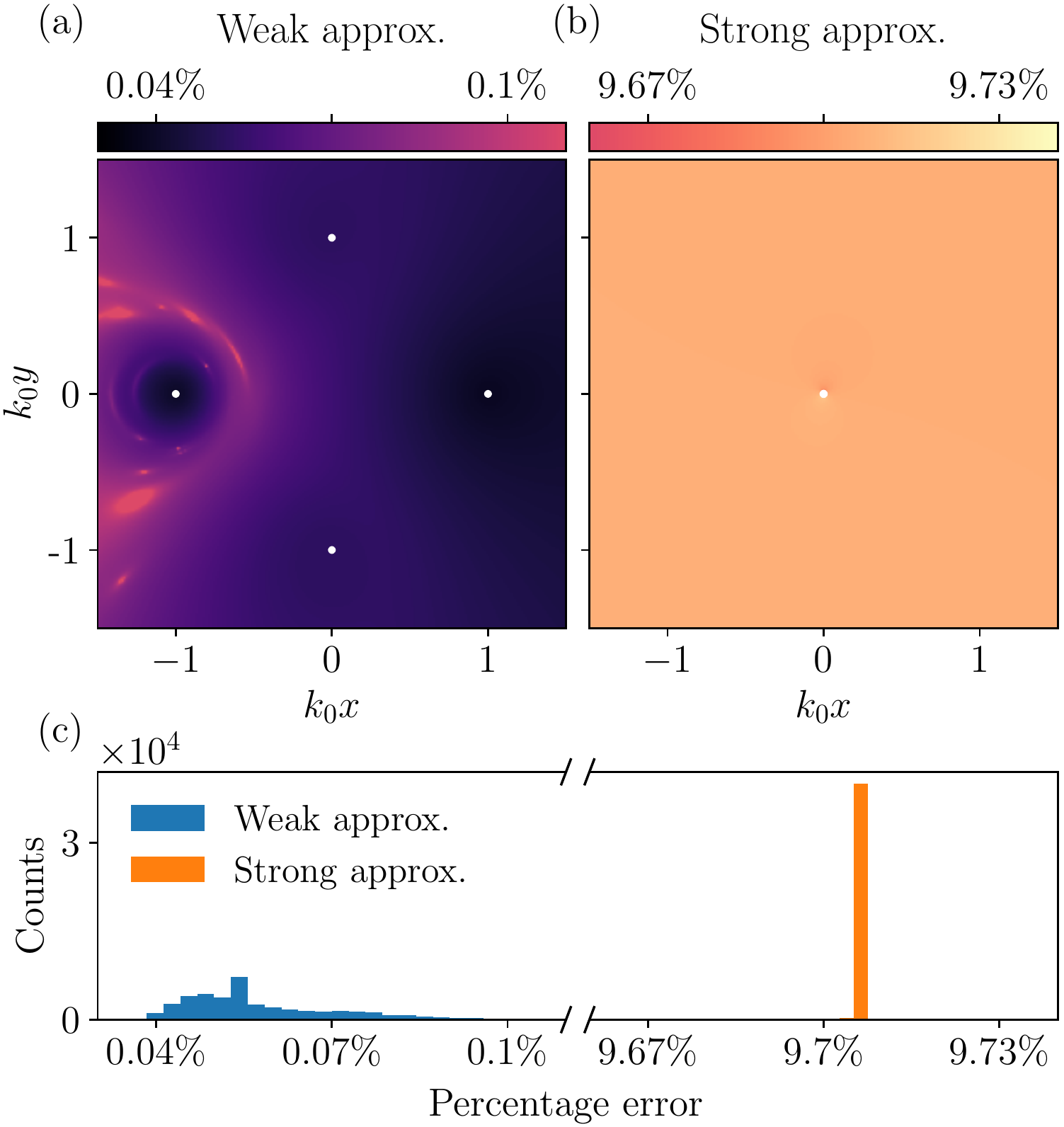}
    \caption{Percentage error of the weak (panel (a)) and strong (panel (b)) coupling approximations of the total field. This is evaluated with respect to the corresponding non--approximated total field obtained using Eq.~\eqref{eq:inversion}. Since the interaction strengths are determined by the Green's functions, the weak and strong approximations only differ in the inter--scatterer distance, while the remaining parameters are kept unchanged. The error is averaged over 100 setups with random polarizabilities. The white dots identify the scatterers in the system. Given the small inter--scatterer distance of the strong coupling approximation, in panel (b), the scatterers are represented all on top of each other. In panel (c), while the strong--coupling approximation (orange) maintains a uniform percentage error in space, the weak--coupling approximation (blue) strongly depends on spatial distribution.}
    \label{fig:approx}
\end{figure}

Unlike the Born series, which typically diverges in the limit of strong scattering, we can also take the limit of very strongly coupled particles, isolating those graphs with the largest number of edges (i.e., the highest non--trivial power of the inter--particle interaction $G$). Thus, we keep only the highest--order interaction terms of the sum in the adjugate terms and in the full determinant. In Fig.~\ref{fig:graphs_approx} panel (b), we see how these correspond to 1-connections of order $N-1$ for the adjugate and linear subdigraphs of order $N$ for the determinant. Consequently, the most significant scattering event in the case of strongly interacting scatterers is represented by a signal traveling across the entire system and interacting with the highest number of scatterers\footnote{Note that, although these approximations select a small subset of all the possible scattering processes, their number still increases rapidly with the number of particles $N$.}. Therefore, graph theory allows for a systematic way to calculate the total field $\phi(\bm{x})$ to any order in the interaction.

This graph interpretation results in a very efficient way of getting a good approximation of the total field $\phi(\bm{x})$ while only including the dominant scattering events in the weak ($0^\mathrm{th},1^\mathrm{st},2^\mathrm{nd}$--order) and strong ($N^\mathrm{th},(N-1)^\mathrm{th}$--order) cases. We show this in Fig.~\ref{fig:approx}, where we evaluate the average percentage error of the absolute value of the approximated fields $|\phi(\bm{x})^{\mathrm{weak}}|$ (in panel (a)) and $|\phi(\bm{x})^{\mathrm{strong}}|$ (in panel (b)) against the absolute value of the corresponding non--approximated field $|\phi(\bm{x})|$. The percentage error is averaged over 100 random values of scatterers' polarizations.


\section{N--th order exceptional points}
\label{sec:nthorderEP}
An exceptional point (EP) of a system is a non--Hermitian degeneracy in parameter space that emerges whenever two or more eigenvectors coalesce. The order of the EP is determined by the number of coalescing eigenvectors. At the EP, the matrix of the system is not diagonalizable but still admits a Jordan form~\cite{Horn_2012}. In such form, the dimension of the Jordan blocks correspond to the order of the eigenvectors' coalescence, e.g., a $2\times2$ Jordan block corresponds to a $2^\mathrm{nd}$--order coalescence and so on. Finding these non--Hermitian singularities in small--dimensional systems is straightforward and an analytical solution can be quickly determined. Both $2^\mathrm{nd}$--order and limited higher--order EPs have been thoroughly studied~\cite{Demange_2011,Wiersig_2014,Lin_2016} and experimentally realized~\cite{Dembowski_2001,Chen_2017,Hodaei_2017}. However, no consistent method to find $N^\mathrm{th}$--order EPs in wave scattering systems has been presented yet. Note that we focus on those EPs with degenerate zero eigenvalue due to their clear physical implications on the total field of the system. In fact, since the total field depends on the inverse of the determinant, these eigenvalues are the cause to its highly degenerate responsiveness to parameter perturbation.

In the following, we use the transpose Frobenius companion matrix and its characteristic polynomial to explore $N^\mathrm{th}$--order zero eigenvalue EPs~\cite{Sayyad_2022} and we interpret the result from a graph theory perspective. Note that, in a similar fashion, companion matrices and $N$-th order EPs have been recently studied in a tropical geometric framework~\cite{banerjee2023tropical}. We then design an EP in a scattering setting and probe the system's response against parameter perturbations.

\subsection{EPs conditions}
\label{ssec:EPconditions}
We now consider a system of $N$ scatterers and impose the condition that, at some desired resonant frequency $\omega_0$, the interaction matrix \eqref{eq:M} exhibits an $N^\mathrm{th}$--order EP whose eigenvalues coalesce to zero. As the outgoing field from the system depends on the inverse of the interaction matrix, this ought to yield a system whose power output diverges at the design frequency, and yet is also very sensitive to small perturbations (as in~\cite{Hodaei_2017}), e.g., the scatterer positions.

We first consider the transpose Frobenius companion matrix $\bm{M}_\mathrm{Frob}$ associated with the matrix $\bm{M}$ of Eq.~\eqref{eq:M}~\cite{Brand_1964}. The companion matrix is defined such that it generates the same polynomial for the eigenvalues $\lambda$ of $\bm{M}$, and is given by
\begin{equation}\label{eq:matrix_frobenius}
    \bm{M}_\mathrm{Frob} =
    \begin{pmatrix}
        0 & 1 & 0 & \cdots & 0 & 0\\
        0 & 0 & 1 & \cdots & 0 & 0\\
        \vdots & \vdots & \vdots & \ddots & \vdots & \vdots \\
        0 & 0 & 0 & \cdots & 1 & 0 \\
        0 & 0 & 0 & \cdots & 0 & 1 \\
        -c_0 & -c_1 & -c_2 & \cdots & -c_{N-2} & -c_{N-1}
    \end{pmatrix} ,
\end{equation}
where the $c_i$ are the coefficients of the powers of $\lambda$ in the characteristic polynomial,
\begin{align}
    \label{eq:polynomial}
    0 &= \det(\lambda \mathds{1}-\bm{M}) \\[3pt]\nonumber
    &= \det(\lambda \mathds{1}-\bm{M}_{\mathrm{Frob}}) \\[3pt]\nonumber
    &= \lambda^N+(-1)^{1}c_{N-1}\lambda^{N-1}+\cdots+(-1)^{N}c_1\lambda+c_0 .
\end{align}
The form of the companion matrix is useful to us as it is closely related to the single $N\times N$ Jordan block matrix, $\bm{J}=\delta_{i+1,j}$ where $i,j \in [1, N]$,
\begin{equation}\label{eq:matrix_jordan}
    \bm{J} =
    \begin{pmatrix}
        0 & 1 & 0 & \cdots & 0 & 0\\[2pt]
        0 & 0 & 1 & \cdots & 0 & 0\\[2pt]
        \;\vdots\; & \;\vdots\; & \;\vdots\; & \;\ddots\; & \;\vdots\; & \;\vdots\; \\[2pt]
        0 & 0 & 0 & \cdots & 1 & 0 \\[2pt]
        0 & 0 & 0 & \cdots & 0 & 1 \\[2pt]
        0 & 0 & 0 & \cdots & 0 & 0
    \end{pmatrix} .
\end{equation}
The two matrices \eqref{eq:matrix_frobenius} and \eqref{eq:matrix_jordan} take the same form once all the $c_i$ in \eqref{eq:matrix_frobenius} are zero. We assume that the interaction matrix $\bm{M}$ in Eq.~\eqref{eq:M} differs from \eqref{eq:matrix_frobenius} by a similarity transformation, an assumption which holds for the cases considered below. It is, in fact, sufficient for the interaction matrix $\bm{M}$ to have $N$ distinct roots (in regime of no EPs) for the transformation $\bm{M}_\mathrm{Frob}=\bm{T}^{-1}\bm{M}\bm{T}$ to exist~\cite{Ding_2010}. The transformation matrix $\bm{T}=\bm{P}\bm{Q}^{-1}$ is derived as the product of the non-singular matrix $\bm{P}$ whose columns are the eigenvectors of $\bm{M}$ and $\bm{Q}$ whose columns are made of the set of $N$ eigenvectors of $\bm{M}_\mathrm{Frob}$, $\bm{q}_i=(1, \lambda_i, \lambda_i^2, \cdots, \lambda_i^{N-1})^T$ relative to its eigenvalues $\lambda_i$~\cite{Bellman_1997}. \footnote{Note that, in case there is no similarity transformation between the matrix and its Frobenius companion matrix, it is always possible to find lower order EPs given by the block companion matrices Ref.~\cite{Ding_2010}.} For details on the derivation of such transformation $\bm{T}$ see App.~\ref{app:transformation}. With this assumption, there is an $N^\mathrm{th}$--order non--Hermitian degeneracy in the spectrum of $\bm{M}$ when all the $c_i$ are zero. By means of this simple requirement, we can engineer an zero eigenvalue EP of desired order by solving the set of non--linear equations given by the conditions $c_i=0$ for $i=0,1,\cdots,N-1$. These coefficients $c_i$ can be evaluated relying on the expansion of the determinant in terms of its minors. Our system of equations for an $N^\mathrm{th}$--order EP with zero eigenvalue thus becomes
\begin{gather}\label{eq:nonlinearsystem}
    \left\{
    \begin{aligned}
        c_0 &= \det(\bm{M}) = 0 \\[3pt]
        c_1 &= \sum_{I_1\in\mathcal{S}_1([n])} \det(\bm{M}_{(i_1,i_1)}) = 0 \\[3pt]
        c_2 &= \sum_{I_2\in\mathcal{S}_2([n])} \det(\bm{M}_{(i_1,i_1),(i_2,i_2)}) = 0 \\[3pt]
        \;\vdots\;& \\[3pt]
        c_{N-1} &= \sum_{I_{N-1}\in\mathcal{S}_{N-1}([n])}\det(\bm{M}_{(i_1,i_1),\cdots,(i_{N-1},i_{N-1})}) \\[3pt] &= \Tr(\bm{M}) = 0 ,
    \end{aligned}
    \right.
    \raisetag{\baselineskip}
\end{gather}
where $I_m$ is the set of indices $I_m = \{i_1, i_2, \cdots, i_m\}$ defining the minor and $\mathcal{S}_m([n])$ is the collection of size--$m$ combinations within the set $[n] = \{1,2,\cdots,n\}$. Therefore, $\displaystyle\bm{M}_{(i,i)}$ is the first minor obtained by removing the $i$--th row and column, $\displaystyle\bm{M}_{(i,i),(j,j)}$ is the second minor obtained by removing $i$--th and $j$--th rows and columns, and so on. Using this form to construct the coefficients $c_i$, we numerically evaluate the solution to the non--linear system, identifying the parameters for an $N^\mathrm{th}$--order EP.

Importantly, the EP conditions \eqref{eq:nonlinearsystem} are given in terms of sums of minors of the interaction matrix, which we have given a graph theoretic interpretation for in Eq.~\eqref{eq:det} and Eq.~\eqref{eq:adj}. For instance, satisfying the final condition in Eq.~\eqref{eq:nonlinearsystem} requires a vanishing sum of the $1\times1$ minors, which equals the trace of the interaction matrix. From the identification shown in Fig.~\ref{fig:k4}, this condition requires the vanishing sum of the self--interactions in the system. Thus, at a zero eigenvalue $N^\mathrm{th}$--order exceptional point, we require (among others) the condition that the inverse polarizabilities $\alpha_{i}^{-1}$ sum to zero. Since the polarizabilities are complex, both the real and imaginary parts of the $\alpha$ will have to sum to zero, which is only possible in the presence of active scatterers, i.e., scatterers that exhibit gain. Moving up through the conditions \eqref{eq:nonlinearsystem}, from $c_{N-1}$ to $c_{N-2}$ and so on, we see that all the second order interactions within the $2\times2$  minors must also sum to zero (equivalent to considering the $2\times2$ interaction matrix for every pair of particles in the system), as must the third order interactions defined within the $3\times3$ minors and so on. We thus reach the conclusion that an $N^\mathrm{th}$--order exceptional point can be associated with $N$ conditions, each requiring the vanishing sum of sub--scattering events between a fixed number of particles. Note that the latter zero trace and determinant conditions found in the scattering matrix are reminiscent of the ones found in the case of systems described by a Hamiltonian with pseudochiral symmetry \cite{Sayyad_2022}.

In addition to the maximal $N^\mathrm{th}$--order EP, we can also find $n^\mathrm{th}$--order singularities with $n<N$ by requiring only the first $n$ coefficients $c_0$, $c_1$, $\cdots$, $c_{n-1}$ to vanish. This generates a smaller non--linear system whose solution identifies an $n^\mathrm{th}$--order EP. This is only possible if $n$ coefficients vanish in ascending order, starting from $c_0$. In fact, this condition allows one to collect a factor $\lambda^n$ in the polynomial in Eq.~\eqref{eq:polynomial}, producing an $n^\mathrm{th}$--order $\lambda=0$ solution. This solution corresponds to the $n\times n$ Jordan block relative to the $n^\mathrm{th}$--order EP. Any other combination of vanishing coefficients results in a diagonalizable system, without non--Hermitian singularities. Finally, note that the construction of EPs is inevitably dependent on the presence of interaction $G$ in the system. In fact, in the case of no interaction, we would be left with a diagonalizable system.


\subsection{Graph theory conditions for EPs}
\label{ssec:graphEP}
As an example, we now design a scattering configuration exhibiting a $4^\mathrm{th}$--order exceptional point and we interpret the condition of non-Hermitian degeneracy in terms of graphs. In the next subsection, we show how the scattered total field depends on a chosen parameter, in our case, the position of the first scatterer.

\begin{figure}
    \centering
    \includegraphics[width=0.5\linewidth]{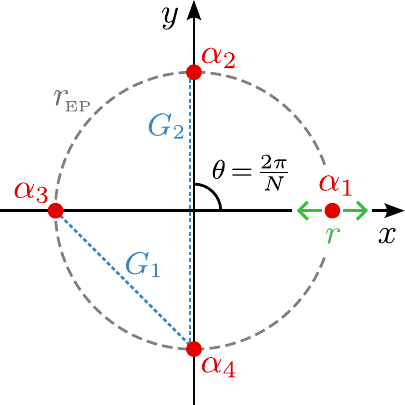}
    \caption{Example construction of a scattering system with an $N^\mathrm{th}$--order exceptional point. The system consists of $N$ scatterers (here $N=4$) with polarizabilities $\alpha_n$ forming a cyclic polygon on a circle with radius $r_\mathrm{EP}$. Since the scatterers are equidistantly spaced, the angle $\theta$ is uniquely determined by the number of scatterers $N$, $\theta=2\pi/N$. The scatterers interact with the nearest neighbors via the Green's function $G_1$ and with the next--to--nearest neighbors via $G_2$. When probing the total field and the power output, we use the radial distance of the first scatterer $r$ as the tunable parameter to scan through the exceptional point in parameter space.}
    \label{fig:system}
\end{figure}
\begin{figure}
    \centering
    \includegraphics[width=\linewidth]{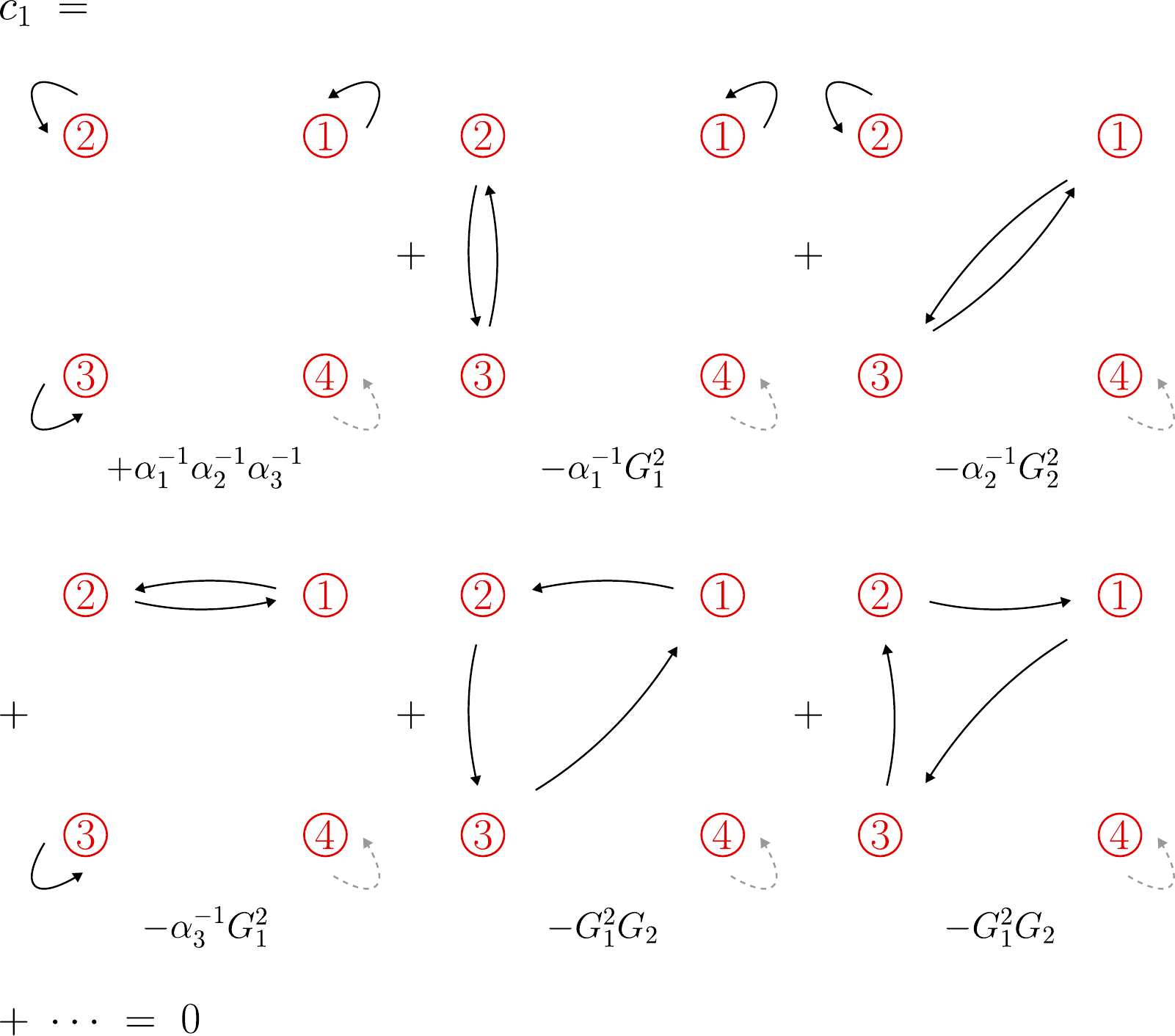}
    \caption{Example construction of the condition $c_1=0$ represented by the appropriate 1--connection graphs. The conditions are the graph--theory analogous of the set of non--linear equations in \eqref{eq:nonlinearsystem} for the interaction matrix $\bm{M}_\mathrm{4,sym}$ relative to Fig.~\ref{fig:system}. We show only the events $D^*[4\rightarrow 4]$ with the $4^\mathrm{th}$ scatterer neglected, however the condition accounts also for three analogous sets of graphs in which the other scatterers are neglected, namely $D^*[1\rightarrow 1]$, $D^*[2\rightarrow 2]$, and $D^*[3\rightarrow 3]$. All the resulting scattering events have to be finally summed together to give the final condition $c_1=\alpha_1^{-1}\alpha_2^{-1}\alpha_3^{-1}-\alpha_1^{-1}G_1^2-\alpha_2^{-1}G_2^2-\alpha_3^{-1}G_1^2-2G_1^2G_2+\cdots=0$.}
    \label{fig:c1condition}
\end{figure}

\begin{figure*}
    \centering
    \includegraphics[width=0.82\linewidth]{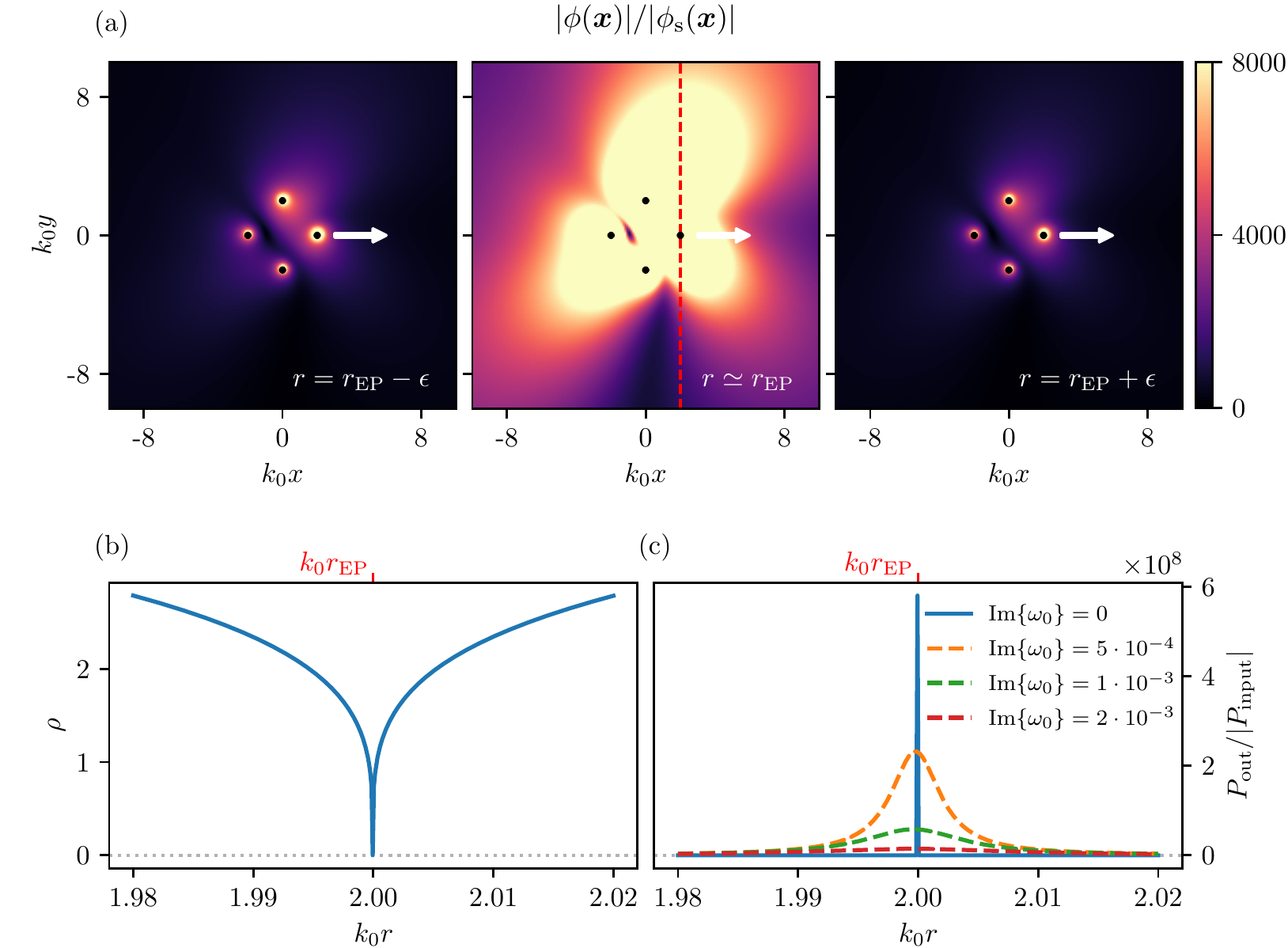}
    \caption{Effects of a $4^\mathrm{th}$--order exceptional point on the total field (panel (a)), global Euclidean distance (panel (b)), and power output (panel (c)) of the system in response to a change in the tuning parameter, that is, the radial distance of the first scatterer $r$. The latter ranges in $r\in[r_\mathrm{EP}-\epsilon k_0^{-1}, r_\mathrm{EP}+\epsilon k_0^{-1}]$, where $\epsilon$ defines a small deviation from the exceptional point. In panel (a), we show the absolute value of the total field normalized against the source field. In the scan from left to right (indicated by the white arrow), the tunable radial distance $r$ is shifted by the amounts $\epsilon\in[-10^{-3}, -10^{-5}, +10^{-3}]$. Note how the total field experiences a sudden peak in the proximity of the EP (middle plot). In panel (b), we show the global Euclidean distance of the right eigenvectors (see Eq.~\eqref{eq:eucldist}) as a function of the tuning parameter $r$. This measure goes to zero when $r=r_\mathrm{EP}$. At this point, all the right eigenvectors (and corresponding left eigenvectors) merge into a single one. In panel (c), we show the power output (see Eq.~\eqref{eq:power}) with respect to the tunable parameter $r$ for different purely imaginary shifts of the resonant frequency, $\Im{\omega_0}\in\{0, 5\cdot10^{-4}, 1\cdot10^{-3}, 2\cdot10^{-3}\}$. For increasing imaginary shifts, the power response of the system broadens while the peak power at the EP reduces. Note that, given the general high gain of the system determined by the polarizabilities, the baseline power output remains of order $10^{7}$ even for significant shifts from the ideal EP condition.}
    \label{fig:ep}
\end{figure*}

For the purpose of simplicity and readability, we now find the parameters (in our case, the polarizabilities $\alpha$) that satisfy the EP conditions in a system in which the scatterers' positions are fixed. As sketched in Fig.~\ref{fig:system}, we equidistantly inscribe our scatterer array in a circle of radius $r_\mathrm{EP}$, simplifying the interaction matrix such that it contains only $N/2$ different Green's functions $G$ when $N$ is even, and $(N-1)/2$ when $N$ is odd. Given the limited number of Green's functions, this configuration is particularly convenient for an efficient search of the EPs. The interaction matrix associated with these cyclic polygons of scattering particles is,
\begin{equation}\label{eq:M_inscribed}
    \bm{M}_\mathrm{sym}=
    \begin{pmatrix}
        \alpha_{1}^{-1} & -G_1 & -G_2 & \cdots & -G_2 & -G_1\\
        -G_1 & \alpha_{2}^{-1} & -G_1 & \cdots & -G_3 & -G_2\\
        -G_2 & -G_1 & \alpha_{3}^{-1} & \cdots & -G_4 & -G_3\\
        \vdots & \vdots & \vdots & \ddots & \vdots & \vdots \\
        -G_2 & -G_3 & -G_4 & \cdots & \alpha_{N-1}^{-1} & -G_1 \\
        -G_1 & -G_2 & -G_3 & \cdots & -G_1 & \alpha_{N}^{-1}
    \end{pmatrix} ,
\end{equation}
where $G_1$ represents the nearest--neighbor interactions, $G_2$ represents the next--to--nearest--neighbor interactions, and so on. The angle between two consecutive scatterers is $\theta=2\pi/N$. In the figure, we also represent the tunable parameter, that is, the radial distance of the first scatterer $r$. While this parameter is not used to find the EP condition of Eq.~\eqref{eq:M_inscribed} (it would indeed change the periodic--chain--like structure of the matrix in Eq.~\eqref{eq:M_inscribed}), it will be needed later for the numerical analysis on the system's sensitivity to parameter perturbations.

Our system is described by the $4\times 4$ matrix $\bm{M}_\mathrm{4,sym}$ with $G_1=G(\bm{x}_1,\bm{x}_2)=G(\bm{x}_1,\bm{x}_4)=G(\bm{x}_2,\bm{x}_3)=G(\bm{x}_3,\bm{x}_4)$ and $G_2=G(\bm{x}_1,\bm{x}_3)=G(\bm{x}_2,\bm{x}_4)$. The Frobenius companion matrix of $\bm{M}_\mathrm{4,sym}$ takes the form of Eq.~\eqref{eq:matrix_frobenius} restricted to the space of $4\times 4$ matrices, therefore including only the coefficients $c_i$ with $i \in \{0,1,2,3\}$. These coefficients can be evaluated using the determinants in Eq.~\eqref{eq:nonlinearsystem}.

Our graph theory description previously introduced illustrates the meaning of this set of vanishing sums. For example, in Fig.~\ref{fig:c1condition}, we show the condition $c_1=0$ which requires all the $3^\mathrm{rd}$--order scattering events to sum to zero. It is worth recalling that the zero condition of the $i^\mathrm{th}$--order coefficient is entirely independent of scattering events of any other order. This means that asking for the single coefficient $c_{i}$ to be zero is equivalent to asking for all the scattering events of order $N-i$ to sum to zero. Thus, to find a $4^\mathrm{th}$--order EP, we need the condition $c_i=0$ to be satisfied by the scattering events of \emph{every} order, that is, $c_i=0$ for $i = 0,1,2,3$.

We finally note that, while a graph can be associated to the matrix of eigenvectors of the system, we could not find any particular interpretation to the coalescence of multiple eigenvectors in terms of graphs. Moreover, in the case of EPs of non--trivial order, a mathematical expression for the eigenvectors becomes highly cumbersome and strongly dependent on the system described. The non--trivial problem of finding a general expression for the eigenvectors of high--order EPs and an associated graph theoretic interpretation is left for further studies.


\subsection{Trading sensitivity for dissipation balance}
\label{ssec:trading}
In Sec.~\ref{ssec:graphEP}, we gave an example of a convenient system to find a $4^\mathrm{th}$--order EP. On this system, we interpreted the condition to find such EPs from a graph theory perspective. We now show how the presence of this high--order EP affects the total field of the system with respect to perturbations to the chosen parameter. In our case, this parameter is the position of the first scatterer $r$ as depicted in Fig.~\ref{fig:system}.

In Fig.~\ref{fig:ep} panel (b), we show the coalescence of the eigenvectors in the range of parameter $r\in[r_\mathrm{EP}-\epsilon k_0^{-1}, r_\mathrm{EP}+\epsilon k_0^{-1}]$ with $\epsilon=0.02$ by means of the vanishing total Euclidean distance. This distance is defined as
\begin{equation}\label{eq:eucldist}
    \rho\coloneqq\sum_{\substack{i=1\\j=i+1}}^N ||\bm{v}_i-\bm{v}_j||,
\end{equation}
where $\bm{v}_i$ and $\bm{v}_j$ are the right eigenvectors of the matrix $\bm{M}_\mathrm{4,sym}$ and the sum takes care of not double--counting terms. This quantity vanishes when $r=r_\mathrm{EP}$, signaling the coalescence of all the $N$ eigenvectors relative to the degenerate eigenvalue $0$. This is the $N^\mathrm{th}$--order exceptional point. Note that, given the high--order nature of the exceptional point, known EP measures like the phase rigidity of the eigenvectors and the condition number of the eigenvector matrix do not entirely capture the features of the singularity~\cite{Scali_2021}. Note also that while the distance in Eq.~\ref{eq:eucldist} serves as an intuitive quantity to witness full eigenvector degeneracy, it is unable to give insight on the eigenvector scaling around the EPs. To do so, one can still access the phase rigidity's critical exponent~\cite{rotter2010role,Tang_2020}. The immediate effects of the EP on the total field are shown in Fig.~\ref{fig:ep} panel (a). In this figure, we scan, from left to right, through the EP with the tunable parameter $r$. In proximity of the EP, the absolute value of the total field $|\phi(\bm{x})|$ rapidly increases before attenuating again, once the singularity is passed.

In the same way, we can probe the EP just obtained by measuring the power output of our system of scatterers (see Fig.~\ref{fig:system}), which we define as
\begin{align}\label{eq:power}
    P_{\mathrm{out}} \coloneqq &\oint_\mathrm{S}\Im{\phi(\bm{x})^*\nabla\phi(\bm{x})}\cdot \hat{n}\;\dd s \\ \nonumber
    = &-\sum_{\substack{n=1\\n\neq m}}^N \Im{\alpha_n}|\phi_\mathrm{inc}(\bm{x}_n)|^2 .
\end{align}
Eq.~\eqref{eq:power} is derived, after little manipulation, by integrating the LHS of Eq.~\eqref{eq:helmholtz} (multiplied from the left by the complex conjugate field $\phi(\bm{x})^*$) in a volume surrounding all the scatterers. We obtain the closed surface integral in Eq.~\eqref{eq:power} by means of the divergence theorem.

The power output, as written in Eq.~\eqref{eq:power}, depends on the sum of the incident fields on the different scatterers of the system weighted by the imaginary parts of the polarizabilities. In our case, the entire dependence of the power response on the tuning parameter $r$ is contained in the incident field. This is uniquely determined by the matrix $\bm{M}_\mathrm{sym}$. It is common to express the sensitivity (in our case, in the form of power output) of the system at the EPs in terms of a perturbation to the system matrix~\cite{Gunther_2007,Hodaei_2017}. Thus, to express the power output in Puiseux series, one would need to rederive the scattering matrix in terms of a perturbation around the EP as for example $\bm{M}_\mathrm{sym}=\bm{J}+\varepsilon\bm{M}'$ where $\bm{J}$ is the full Jordan matrix~\eqref{eq:matrix_jordan} and $\bm{M}'$ is a non trivial perturbation matrix~\cite{Sayyad_2022}. Doing so, if the perturbation around the EP lifts the coefficient $c_{N-1}$ such that $c_{N-1}\neq 0$, the Puiseux series $\lambda = \lambda_0 + \sum_{i=1}^\infty\varepsilon^{i/N}\lambda_i$ exists and refers to the $N^\mathrm{th}$--order EP. However, in case the perturbation leaves $c_{N-1}=0$, the perturbed eigenvalues split in $k$ different cycles of order $n_k$ of the form $\lambda_k = \lambda_0 + \sum_{i=1}^\infty \varepsilon^{i/n_k}\lambda_{k,i}$ with the various $n_k<N$ summing to $N$ as $\sum_k n_k = N$~\cite{Ma_1998,Demange_2011,Sayyad_2022}. Note that, in the case of the system described in Eq.~\eqref{eq:M_inscribed}, a perturbation in the radial distance $r$ indeed lifts the coefficient $c_{N-1}$ such that $c_{N-1}\neq 0$.

Given the high order of the exceptional point, the power output of the system shows extreme sensitivity to perturbations in parameter space. In Fig.~\ref{fig:ep} panel (c), we show the power output of Eq.~\eqref{eq:power} versus the tunable parameter $r$ for different imaginary offsets of the resonant frequency $\omega_0$ at which the EP is found.

The introduction of an imaginary part in the design frequency has multiple functions. On the one hand, it helps to understand how possible experimental inaccuracies can affect peak and shape of the power output of the system. On the other hand, it shows how ``ad--hoc'' imaginary shifts in the design frequency of the system can help to adjust the distribution of gain/loss across the scatterers. Since the system is then probed with real frequencies $\omega_0$, introducing an imaginary shift in the design frequency results in a quasi--coalescence of the eigenvectors causing a drop in the system responsiveness to the singularity. This is shown in the figure by means of the amplitude reduction and broadening of the power output curves when increasing the imaginary shift of $\omega_0$. Note that the curve with $\Im{\omega_0}=0$ (solid blue curve in the figure), which is set to cross the EP, is re--scaled by a factor $10^{-26}$ in order to fit into the graph and give some insight of the power output behavior.

\begin{figure}
    \centering
    \includegraphics[width=0.9\linewidth]{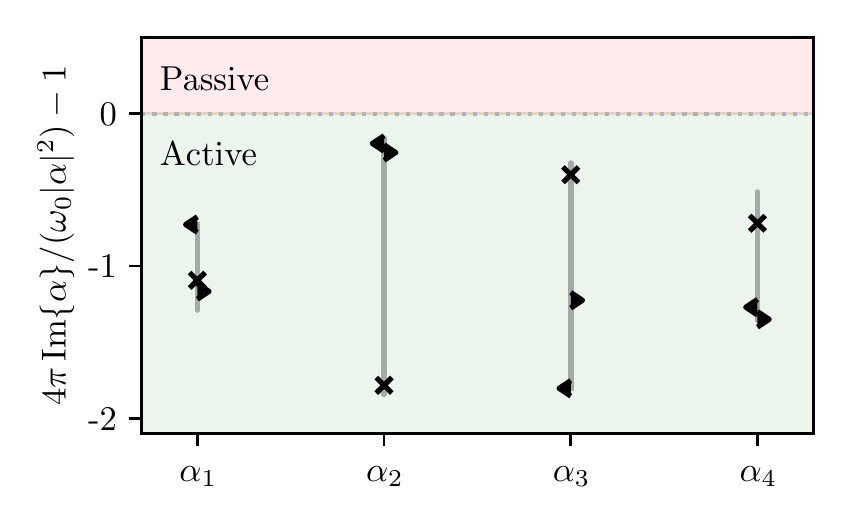}
    \caption{Results of the inequality in Eq.~\eqref{eq:inequality} for the polarizabilities $\alpha$. The polarizabilities are the solutions to the EP conditions \eqref{eq:nonlinearsystem} of the system described by $\bm{M}_\mathrm{4,sym}$ and shown in Fig.~\ref{fig:system}. The solid lines are all the possible solutions of polarizabilities for a scan in the imaginary part of the resonant frequency, $\Im{\omega_0}\in[-1,1]$. This has been done in a similar fashion to Ref.~\cite{Cho_2020}. The ``cross'' marker indicates $\Im{\omega_0}=0$ while the ``left--caret'' and ``right--caret'' indicate the end of the imaginary ranges, $\Im{\omega_0}=-1$ and $\Im{\omega_0}=1$, respectively. The scattering elements are passive when the $\alpha$ lay on the positive semi--plane (red semi--plane), therefore they satisfy the inequality. In the figure, all the solutions $\alpha$ of the system considered are active, therefore laying on the negative semi--plane (green semi--plane). The figure shows how imaginary shifts in the resonant frequency $\omega_0$ used to design the EP allows one to tune the distribution of the gain/loss of the system across the scatterers.}
    \label{fig:inequality}
\end{figure}

We now show how we can tune the distribution of the gain/loss of the system across the scatterers in order to finely adjust possible experimental setups, where it is preferred to have a set of scatterers with the least possible gain. The presence of exceptional points inevitably depends on the scatterers' structure and, in particular, on their active nature. The condition for a scatterer $j$ to be passive is that its polarizability $\alpha_j$ satisfies the inequality~\cite{Markel_2019}
\begin{equation}
    \label{eq:inequality}
    \Im{\alpha_j} > \frac{k_0}{4\pi}|\alpha_j|^2 ,
\end{equation}
obtained by asking for a negative divergence of the power output in the case of passive scatterers. Eq.~\eqref{eq:inequality} is derived for the case of 3D Green's function as considered in this paper. As a reminder, $k_0=\omega_0/c$ with $c=1$ in this paper. If the polarizabilty of a scatterer satisfies this inequality, the scatterer acts as a passive, lossy medium. As we have seen in Eq.~\eqref{eq:nonlinearsystem}, one requirement to obtain an $N^\mathrm{th}$--order EP is $\Tr(\bm{M})=0$, i.e., all the scattering events of 0--th order have to sum to 0 while individually being non--vanishing. This implies having active elements in the system, i.e., scatterers with $\Im{\alpha}<0$ which cannot satisfy Eq.~\eqref{eq:inequality}. On the other hand, elements with $\Im{\alpha}>0$ do not necessarily satisfy Eq.~\eqref{eq:inequality}, thus, are not necessarily passive. By means of this inequality, we define a polarizability regime in which energy has to be injected into the system to obtain these $N^\mathrm{th}$--order EPs.

In Fig.~\ref{fig:inequality}, we show this inequality test for the polarizabilities of the system described by $\bm{M}_{4,\mathrm{sym}}$. In this case, none of the polarizabilities satisfy the inequality (no polarizabilities lie on the positive half of the plane), indicating that no passive scatterers are found in the system \footnote{Using our numerics, we found polarizabilities satisfying the inequality \eqref{eq:inequality} in 7--scatterer systems described by the matrix \eqref{eq:M_inscribed}.}. The test consists of a scan in the imaginary shift range $\Im{\omega_0}\in[-1,1]$, where $\omega_0$ is the resonant frequency at which the EP is evaluated. The ``cross'' marker indicates $\Im{\omega_0}=0$ while the ``left--caret'' and ``right--caret'' indicate the end of the imaginary ranges, $\Im{\omega_0}=-1$ and $\Im{\omega_0}=1$, respectively. The semi--transparent lines represent all the intermediate $\alpha$'s solutions found in this range.

Note that, we already implemented this imaginary--shifted resonant frequency in order to control the spectral width of the scattering resonance of the system (see panel (c) of Fig.~\ref{fig:ep}). However, in this case, one can use the imaginary shift to move the gain/loss bias on different scatterers. Therefore, an imaginary shift in the design resonant frequency allows one to fine tune the dissipation balance of the system in exchange of a broadening of the power output with respect to the EP parameter. This fine tuning capability becomes crucial in experimental setups that aim for the least possible gain in their set of scatterers.


\section{Conclusion}
\label{sec:conclusion}
In this paper, we used graph theory to solve wave scattering problems within the discrete dipole approximation (DDA).

Firstly, we showed how to use graph theory to develop a diagrammatic method for understanding multiple scattering processes. These processes are encoded in the inverse of the interaction matrix used to find the analytical total field of the system. We interpreted single scattering events in terms of 1--connections and linear subdigraphs and used these to approximate weakly and strongly coupled systems. This is a convenient machinery to calculate the total field $\phi(\bm{x})$ when the dimensionality of the system makes finding a full analytical solution impractical.

Secondly, by exploiting the Frobenius companion matrix associated with the system, we developed a systematic procedure to find $N^\mathrm{th}$--order zero eigenvalue exceptional points (EPs). The EPs are found by making vanish the sum of the 1--connections associated with scattering events of the same order. At a zero eigenvalue EP, the scattering becomes singular, causing the divergence of the emitted power. In our example, the perturbation coincided with a single--particle displacement from the EP configuration of the order of $1/100$ of a wavelength. Although such a sharp sensitivity is achieved in position basis, one could describe the system in terms of the directions of input and output waves. Note that, as shown in this paper, one can also generate $n^\mathrm{th}$--order zero eigenvalue EPs where $n<N$. This might be useful to trade part of the scattered field sensitivity with a reduced number of conditions in the non--linear system. This further reduces the requirement for gain, crucial in certain experimental settings. The generation of $N^\mathrm{th}$--order zero eigenvalue EPs can be of particular interest for coherent perfect absorption (CPA) structures~\cite{Sweeney_2019,Wang_2021}. Here, the signature of the zero eigenvalue EPs (referred to as CPA EPs) is a quartic behavior of the absorption line shape in the perfectly absorbed channel. In addition, we believe the graph theoretical approach to be a promising tool to describe EPs associated with PT symmetry breaking in scattering systems~\cite{Krasnok_2019,Krasnok_2021} and the non-Hermitian skin effect in the case of non--reciprocal 1D chains of scatterers~\cite{Ghaemi_Dizicheh_2021,Zhang_2022,Xin_2023}.

Finally, to control the spectral width of the exceptional points, we explored the consequences of displacing the design resonant frequency into the complex plane. We found that it is possible to trade the required gain/loss of the single scatterers with a broadened response. This would allow one to choose the preferred dissipation balance throughout the array of elements at the expenses of a reduction in the power output of the system. It might be possible to explore this trade--off as well as the entirety of multiple scattering physics in programmable metamaterials such as those demonstrated by Cho et al.~\cite{Cho_2020}.

\section*{Software package}
The Julia package developed for solving the wave scattering problems found in this paper is available at \href{https://github.com/mekise/graph-theory-dda}{https://github.com/mekise/graph-theory-dda}. Note that, while the code should be easily readable for the user, it is not documented. Reasonable requests may be addressed to SS.


\acknowledgments{SS thanks Federico Cerisola for stimulating discussions. JA and SARH thank the Royal Society for support. SS is supported by a DTP grant from EPSRC (EP/R513210/1). JA acknowledges funding from EPSRC (EP/R045577/1). SARH acknowledges the Royal Society and TATA for financial support through the grant~URF\textbackslash R\textbackslash 211033.\\

The authors declare no conflicts of interest.}


\appendix

\section{Graph theory fundamentals}
\label{app:graphintro}
``A graph $\mathcal{G}$ is an ordered pair of disjoint sets $(\mathcal{V,\mathcal{E}})$, such that $\mathcal{E}$ is a subset of unordered pairs of $\mathcal{V}$''~\cite{Bollob_s_1979}. The set $\mathcal{V}$ defines the vertices of the graph, i.e., the interacting elements of a structure we consider. The interactions between these elements are defined by the edges in the set $\mathcal{E}$. In the case of interacting discrete scatterers, the set of vertices $\mathcal{V}$ represents the set of scatterers and the set of edges $\mathcal{E}$ correspond to the set of interactions between the scatterers. Note that, in general, these interactions are not symmetric. By means of these fundamental blocks, we can translate every matrix $\bm{M}$ of the form Eq.~\eqref{eq:M} into a graph of the form \ref{fig:system}. The resulting graph will represent the polarizabilities $\alpha$ as self--loops (or self--edges) and the Green's functions $G(\bm{x}_i,\bm{x}_j)$ as edges starting from the vertex $i$ and ending in the vertex $j$. This directed edges, from $i$ to $j$, promote the graph to a directed--graph or digraph. As mentioned in the main text, this graph is the Coates digraph $D^*(\bm{M})$ associated with the matrix $\bm{M}$. Note that the asterisk superscript takes care of the historical definition of the Coates digraph, i.e., the digraph associated with the transpose of the matrix we intend to represent \cite{Coates_1959,Brualdi_2008}. In the main text, we refer to this kind of graphs as \emph{vertex--labeled directed weighted simple graph permitting loops}. ``Vertex--labeled'', as the name suggests, indicates that the scatterers are distinguishable, ``directed'' means that interactions between scatterers are not necessarily symmetric, ``weighted'' indicates a non--unit interaction, ``simple'' indicates the presence of a single directional interaction between edges, while ``permitting loops'' identifies a graph that allows for self--interaction, in our case, the polarizabilities.

\subsection{Linear subdigraphs}
\label{app:subdigraphs}
\begin{figure}
    \centering
    \includegraphics[width=\linewidth]{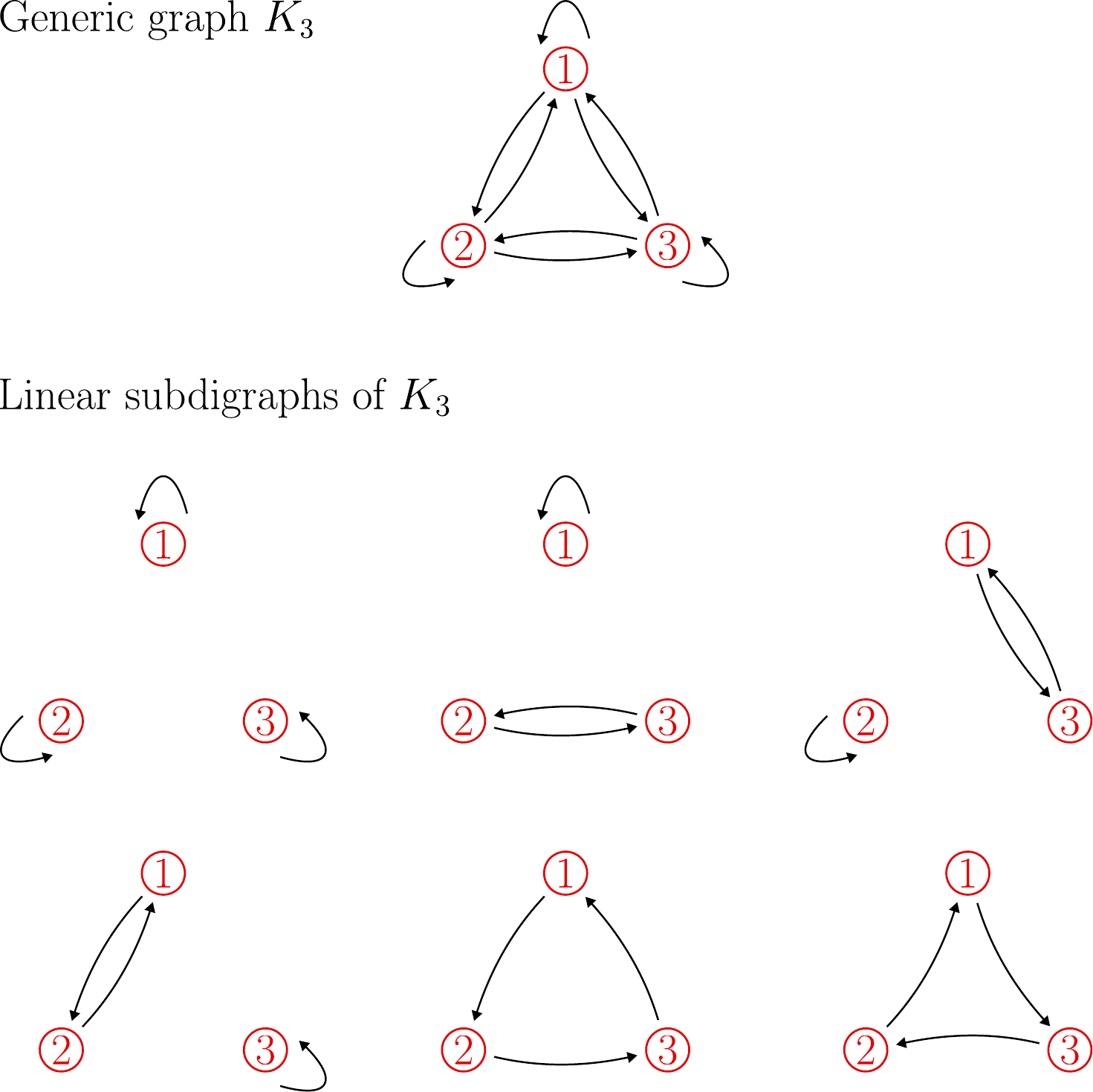}
    \caption{Example of linear subdigraphs associated with the graph $K_3$. For the example graph $K_3$, there are 6 linear subdigraphs in total. Note that, given that we deal with directed graphs, we distinguish between subdigraphs with edges linking the same nodes but in opposite directions as in the case of the last and second--to--last subdigraphs in the figure.}
    \label{fig:subdigraphs}
\end{figure}
\begin{figure}
    \centering
    \includegraphics[width=0.85\linewidth]{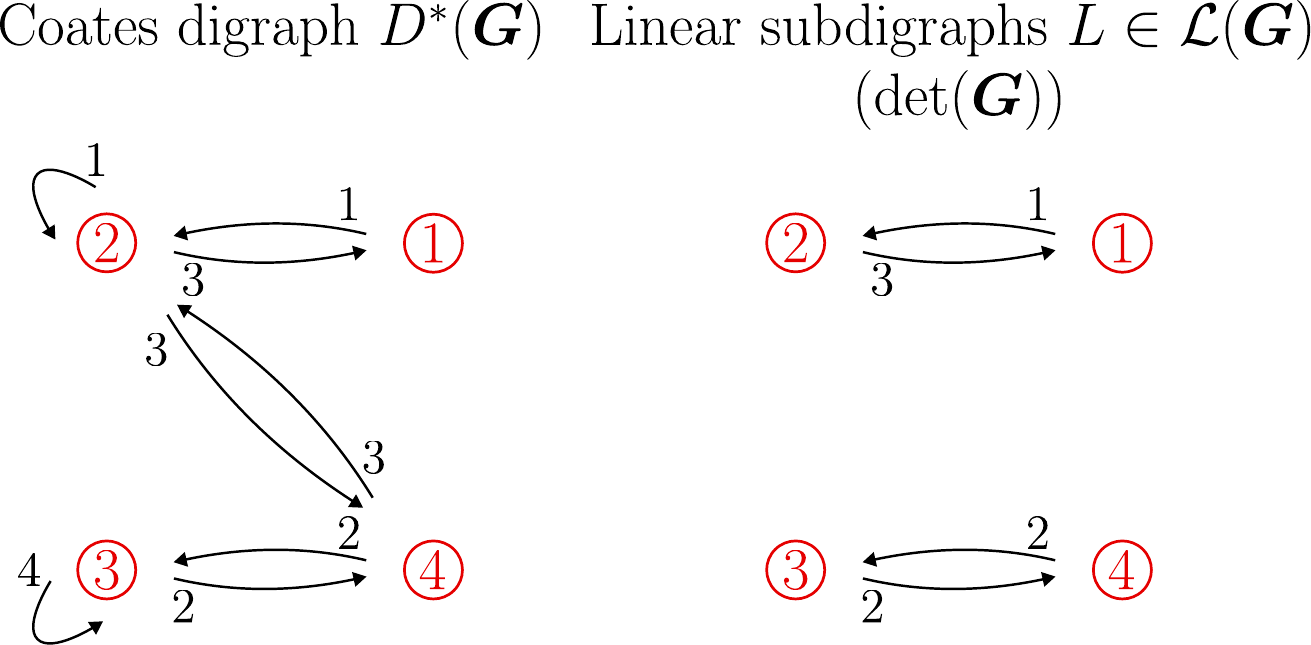}
    \caption{Construction of the determinant of the matrix $\bm{G}$. On the left, the Coates digraph of the matrix $\bm{G}$. On the right, the linear subdigraphs of the matrix $\bm{G}$ which define the determinant as per Eq.~\eqref{eq:det}.}
    \label{fig:det_graph}
\end{figure}
Consider the Coates' determinant formula in Eq.~\eqref{eq:det}, expression for the construction of determinants by means of graphs. We report the expression here for convenience,
\begin{equation}
    \det(\bm{A}) = (-1)^N \sum_{L\in\mathcal{L}(\bm{A})}(-1)^{c(L)}\gamma(L) .
\end{equation}
As a reminder, $N$ is the dimension of a generic matrix $\bm{A}$ whose determinant we want to evaluate, $c(L)$ is the number of cycles in $L$, $\gamma(L)$ is the weight of the linear subdigraph $L$, and $\mathcal{L}(\bm{A})$ is the set of all possible linear subdigraphs of the Coates digraph $D^*(\bm{A})$. We now show what a linear subdigraphs is and how to construct it.

A subdigraph is a digraph with $\mathcal{V}'\subset\mathcal{V}$ vertices and $\mathcal{E}'\subset\mathcal{E}$ edges. In addition, to earn the name of linear subdigraph, the vertices in $V'$ must have in--degree and out--degree equal to 1, i.e., every vertex must have exactly one edge entering and one edge leaving. In Fig.~\ref{fig:subdigraphs}, we report the entire set of linear subdigraphs for an example digraph $\mathcal{K}_3$. In the main text, we use these set of graphs to construct the determinants in the adjugate inversion formula. We now show how we use these linear subdigraph constructions for the determinant evaluation. Consider the sparse matrix $\bm{G}$,
\begin{equation}
    \bm{G}=
    \begin{pmatrix}
        0 & 1 & 0 & 0 \\[3pt]
        3 & 1 & 0 & 3 \\[3pt]
        0 & 0 & 4 & 2 \\[3pt]
        0 & 3 & 2 & 0
    \end{pmatrix} .
\end{equation}
We can work out the digraph associated with the matrix $\bm{G}$ and its linear subdigraphs to evaluate the determinant. To do this, we use Eq.~\eqref{eq:det}, i.e., we search for all the subdigraphs whose vertices have in--degree and out--degree equal to 1. We show the results in Fig.~\ref{fig:det_graph}, where on the LHS we find the digraph $D^*(\bm{G})$ associated with the matrix $\bm{G}$ and on the RHS we find the determinant of $\bm{G}$, consisting of the only linear subdigraph of the graph $D^*(\bm{G})$. Summing the weights of the edges of the subdigraph, we obtain the determinant, $\det(\bm{G})=(-1)^4 (-1)^2 (1\cdot3\cdot2\cdot2)=12$, where the first term accounts for the factor $(-1)^N$, the second term accounts for the number of cycles $(-1)^{c(L)}$, and the last accounts for the weights of the subdigraphs $\gamma(L)$.

\subsection{1--connections}
\label{app:1connection}
\begin{figure}
    \centering
    \includegraphics[width=0.85\linewidth]{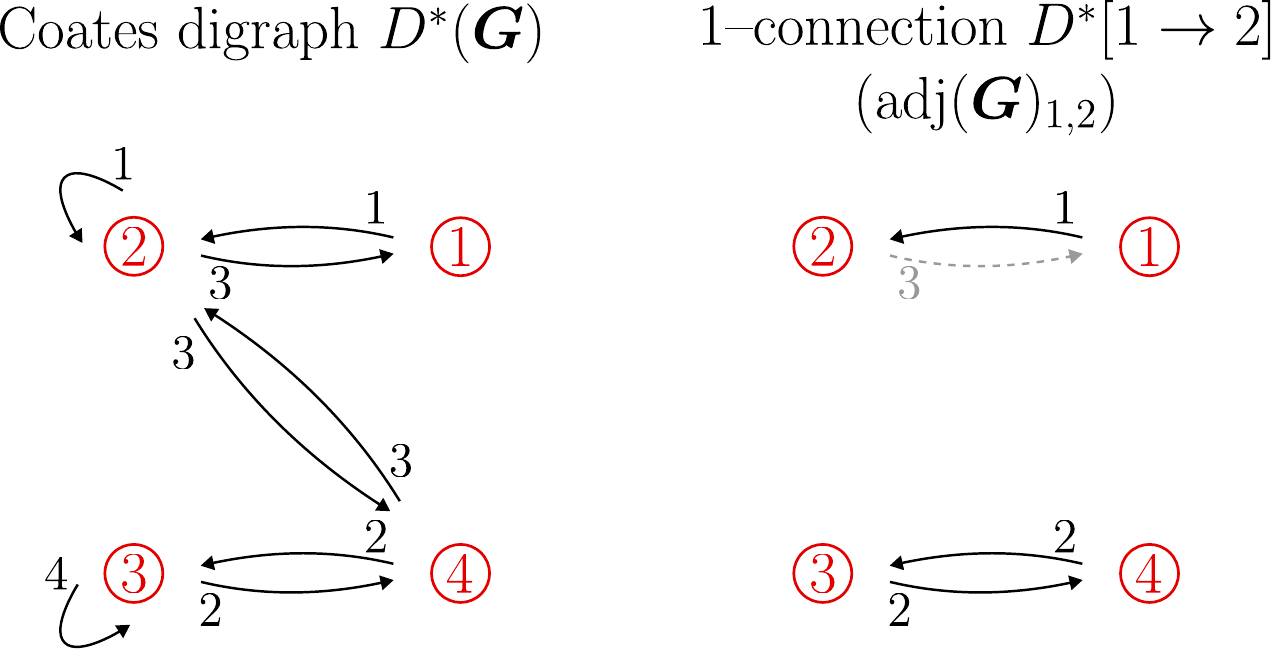}
    \caption{Construction of the adjugate element $\mathrm{adj}(\bm{G})_{1,2}$, built using the off--diagonal 1--connections from vertex 1 to vertex 2. On the left, again the Coates digraph of the matrix $\bm{G}$. On the right, the 1--connections of the matrix $\bm{G}$ which define the adjugate term $\mathrm{adj}(\bm{G})_{1,2}$ as per Eq.~\eqref{eq:adj}. The latter is built by the corresponding linear subdigraphs by removing the edge $2\rightarrow 1$ as described in the text.}
    \label{fig:1connection}
\end{figure}
Consider the adjugate expression in Eq.~\eqref{eq:adj}, expression for the construction of adjugate terms by means of graphs. We report the expression here for convenience,
\begin{equation}
    \mathrm{adj}(\bm{A})_{i,j} = (-1)^N \sum_{D^*[i\rightarrow j]}(-1)^{c(D^*[i\rightarrow j])+1}\gamma(D^*[i\rightarrow j]) .
\end{equation}
In this expression, the terms $D^*[i\rightarrow j]$ are the 1--connections from vertex $i$ to vertex $j$ while all the other elements of the equation have an analogous meaning as in the determinant expression. The 1--connection $D^*[i\rightarrow j]$ is obtained from the corresponding linear subdigraph $L_{\ni i\rightarrow j}$ (linear subdigraph that includes the edge $i\rightarrow j$) by simply removing the edge $j\rightarrow i$. Note that, in the case $i=j$, this corresponds to removing the self--loop at vertex $i$. This definition leads to the following relation between the number of cycles in a linear subdigraph $L_{\ni i\rightarrow j}$ and the relative 1--connection $D^*[i\rightarrow j]$~\cite{Brualdi_2008},
\begin{equation}
    c(L_{\ni i\rightarrow j}) = c(D^*[i\rightarrow j])+1 ,
\end{equation}
which justifies the ``$+1$'' in the adjugate expression. More formally, following the definition of a 1--connection reported in Ref.~\cite{Brualdi_2008}, we call 1--connection from vertex $i$ to vertex $j$, the spanning subdigraph $D^*[i\rightarrow j]$ such that,
\begin{itemize}
    \item for $i\neq j$, all vertices $k$ with $k\neq i,j$ must have in--degree and out--degree equal to 1, vertex $i$ must have in--degree equal to 0 but out--degree equal to 1 and vertex $j$ must have in--degree equal to 1 but out--degree equal to 0. The resulting spanning subdigraph therefore has a path from vertex $i$ to vertex $j$,
    \item for $i=j$, all vertices must have in--degree and out--degree equal to 1, while vertex $i=j$ must have in--degree and out--degree equal to 0.
\end{itemize}
As mentioned in the main text, the 1--connections are closely related to the linear subdigraphs. In fact, the 1--connections $D^*[i\rightarrow j]$ obtained using the definition above are equivalent to those obtained from the corresponding linear subdigraph $L_{\ni i\rightarrow j}$ simply by removing the edge $j\rightarrow i$. By means of this definition, we now show the construction of an off--diagonal adjugate term. Consider again the matrix $\bm{G}$, we now build the adjugate term $\mathrm{adj}(\bm{G})_{1,2}$ consisting of the 1--connections $D^*[1\rightarrow 2]$. To do this, we consider all the linear subdigraphs that include the edge $1\rightarrow 2$ (one single subdigraph in our example) and remove the edge from vertex $2\rightarrow 1$, as shown in Fig.~\ref{fig:1connection}. Summing the weights of the edges of the 1--connections, we obtain the adjugate term, $\mathrm{adj}(\bm{G})_{1,2}=(-1)^4 (-1)^{1+1} (1\cdot2\cdot2)=4$, where the first term accounts for the factor $(-1)^N$, the second term accounts for the number of cycles $(-1)^{c(D^*[i\rightarrow j])+1}$, and the last accounts for the weights of the 1--connections $\gamma(D^*[i\rightarrow j])$.

\section{Similarity transformation between a matrix and its Frobenius companion form}
\label{app:transformation}
Consider a matrix $\bm{A}\in\mathbb{C}^{N\times N}$ and its Frobenius companion matrix (see Eq.~\ref{eq:matrix_frobenius} in the main text),
\begin{equation}
    \bm{A}_\mathrm{Frob} =
    \begin{pmatrix}
        0 & 1 & 0 & \cdots & 0 & 0\\
        0 & 0 & 1 & \cdots & 0 & 0\\
        \vdots & \vdots & \vdots & \ddots & \vdots & \vdots \\
        0 & 0 & 0 & \cdots & 1 & 0 \\
        0 & 0 & 0 & \cdots & 0 & 1 \\
        -c_0 & -c_1 & -c_2 & \cdots & -c_{N-2} & -c_{N-1}
    \end{pmatrix} ,
\end{equation}
where $c_i$ are the coefficients of the characteristic polynomial of $\bm{A}$. If there exists a row vector $\bm{b}\in\mathbb{C}^{1\times N}$ such that the matrix
\begin{equation}
    \bm{T}=
    \begin{pmatrix}
        \bm{b} \\
        \bm{b}\bm{A} \\
        \vdots \\
        \bm{b}\bm{A}^{N-2} \\
        \bm{b}\bm{A}^{N-1} 
    \end{pmatrix}
    \in\mathbb{C}^{N\times N}   
\end{equation}
is non-singular, then the matrix $\bm{A}$ is similar to its Frobenius companion matrix $\bm{A}_\mathrm{Frob}$~\cite{Ding_2010},
\begin{equation}
     \bm{A}_\mathrm{Frob} = \bm{T}^{-1}\bm{A}\bm{T} .
\end{equation}
Here, the matrix $\bm{T}$ is the Vandermonde matrix, whose entries are thus given by
\begin{equation}
    \bm{T}_{ij} = \bm{b}\bm{A}^{i-1}_{:,j},
\end{equation}
where $\bm{A}^{i-1}_{:,j}$ denotes the $j$-th column of $\bm{A}^{i-1}$. Note that, in what follows, we do not compute the vector $\bm{b}$ but rather we infer the transformation matrix $\bm{T}$ while keeping $\bm{b}$ implicit. The Vandermonde determinant can be expressed as
\begin{equation}
    \det(\bm{T}) = \prod_{1\leq i<j\leq N} (\bm{b}\bm{A}^{j-1}-\bm{b}\bm{A}^{i-1}).
\end{equation}
It is immediate to see that $\bm{T}$ is non-singular if and only if $\det(\bm{T}) \neq 0$, thus the $N$ rows $\bm{b}, \bm{b}\bm{A}, \cdots, \bm{b}\bm{A}^{N-2}, \bm{b}\bm{A}^{N-1}$ are distinct. The rows of the Vandermonde matrix are generated by the different powers of $\bm{A}$ multiplied by the same vector $\bm{b}$. Therefore, asking for $\bm{T}$ to be non-singular is equivalent to ask that the characteristic polynomial of $\bm{A}$ has $N$ distinct roots, which requires $\bm{A}$ to be diagonalizable. Thus, we can write $\bm{A}=\bm{P}\bm{D}\bm{P}^{-1}$, where $\bm{D}$ is a diagonal matrix whose entries are the eigenvalues of $\bm{A}$, and $\bm{P}$ is a non-singular matrix whose columns are the eigenvectors of $\bm{A}$. Now, if $\bm{A}$ has $N$ distinct roots, the Frobenius companion matrix can be diagonalized by the matrix $\bm{Q}$ whose columns are made of the set of $N$ eigenvectors of $\bm{A}_\mathrm{Frob}$~\cite{Bellman_1997}
\begin{equation}
    \bm{q}_i=(1, \lambda_i, \lambda_i^2, \cdots, \lambda_i^{N-1})^T
\end{equation}
relative to its eigenvalues $\lambda_i$. Note that a consistent order of the eigenvalues must be kept throughout the calculations. We have that
\begin{equation}
    \bm{A}_\mathrm{Frob} = \bm{T}^{-1}\bm{A}\bm{T}=\bm{T}^{-1}\bm{P}\bm{D}\bm{P}^{-1}\bm{T} = \bm{Q}\bm{D}\bm{Q}^{-1}
\end{equation}
which implies $\bm{Q}=\bm{T}^{-1}\bm{P}$. Since all the matrices in the last expression are invertible, we can derive $\bm{T}=\bm{P}\bm{Q}^{-1}$ and consequently find the transformations $\bm{A}_\mathrm{Frob} = \bm{T}^{-1}\bm{A}\bm{T}$.

As an example, consider the $3\times 3$ matrix
\begin{equation}
    \bm{A} =
    \begin{pmatrix}
        2 & 0 & 0 \\
        -3 & 5 & -4 \\
        -2 & 2 & -1
    \end{pmatrix}
\end{equation}
with eigenvalues $\lambda_1=3, \lambda_2=2, \lambda_3=1$. The similarity matrices $\bm{P}$ and $\bm{Q}$ are,
\begin{equation}
    \bm{P} =
    \begin{pmatrix}
        0 & 1 & 0 \\
        2 & 1 & 1 \\
        1 & 0 & 1
    \end{pmatrix} ,
    \quad
    \bm{Q} = 
    \begin{pmatrix}
        1 & 1 & 1 \\
        3 & 2 & 1 \\
        9 & 4 & 1
    \end{pmatrix} .
\end{equation}
The similarity transformation $\bm{T}$ between the matrix $\bm{A}$ and its Frobenius companion matrix $\bm{A}_\mathrm{Frob}$ thus results
\begin{equation}
    \bm{T} = \bm{P}\bm{Q}^{-1} =
    \begin{pmatrix}
        -3 & 4 & -1 \\
        2 & -3/2 & 1/2 \\
        4 & -4 & 1
    \end{pmatrix} .
\end{equation}
We finally perform the transformation,
\begin{equation}
    \bm{A}_\mathrm{Frob} = \bm{T}^{-1}\bm{A}\bm{T} =
    \begin{pmatrix}
        0 & 1 & 0 \\
        0 & 0 & 1 \\
        6 & -11 & 6
    \end{pmatrix} .
\end{equation}


\bibliographystyle{unsrtnat} 
\bibliography{main}

\begin{thebibliography}{66}
\providecommand{\natexlab}[1]{#1}
\providecommand{\url}[1]{\texttt{#1}}
\expandafter\ifx\csname urlstyle\endcsname\relax
  \providecommand{\doi}[1]{doi: #1}\else
  \providecommand{\doi}{doi: \begingroup \urlstyle{rm}\Url}\fi

\bibitem[Kadic et~al.(2019)Kadic, Milton, van Hecke, and Wegener]{Kadic_2019}
Muamer Kadic, Graeme~W. Milton, Martin van Hecke, and Martin Wegener.
\newblock 3d metamaterials.
\newblock \emph{Nature Reviews Physics}, 1\penalty0 (3):\penalty0 198--210, jan
  2019.
\newblock \doi{10.1038/s42254-018-0018-y}.

\bibitem[Pitruzzello and Krauss(2018)]{Pitruzzello_2018}
Giampaolo Pitruzzello and Thomas~F Krauss.
\newblock Photonic crystal resonances for sensing and imaging.
\newblock \emph{Journal of Optics}, 20\penalty0 (7):\penalty0 073004, jun 2018.
\newblock \doi{10.1088/2040-8986/aac75b}.

\bibitem[Gigan(2022)]{Gigan_2022}
Sylvain Gigan.
\newblock Imaging and computing with disorder.
\newblock \emph{Nature Physics}, 18\penalty0 (9):\penalty0 980--985, sep 2022.
\newblock \doi{10.1038/s41567-022-01681-1}.

\bibitem[Cao(2003)]{Cao_2003}
Hui Cao.
\newblock Lasing in random media.
\newblock \emph{Waves in Random Media}, 13\penalty0 (3):\penalty0 R1--R39, jul
  2003.
\newblock \doi{10.1088/0959-7174/13/3/201}.

\bibitem[Bender and Boettcher(1998)]{Bender_1998}
Carl~M. Bender and Stefan Boettcher.
\newblock Real spectra in non-hermitian hamiltonians having pt symmetry.
\newblock \emph{Physical Review Letters}, 80\penalty0 (24):\penalty0
  5243--5246, jun 1998.
\newblock \doi{10.1103/physrevlett.80.5243}.

\bibitem[Horsley et~al.(2015)Horsley, Artoni, and Rocca]{Horsley_2015}
S.~A.~R. Horsley, M.~Artoni, and G.~C.~La Rocca.
\newblock Spatial kramers--kronig relations and the reflection of waves.
\newblock \emph{Nature Photonics}, 9\penalty0 (7):\penalty0 436--439, jun 2015.
\newblock \doi{10.1038/nphoton.2015.106}.

\bibitem[Makris et~al.(2015)Makris, Musslimani, Christodoulides, and
  Rotter]{Makris_2015}
K.~G. Makris, Z.~H. Musslimani, D.~N. Christodoulides, and S.~Rotter.
\newblock Constant-intensity waves and their modulation instability in
  non-hermitian potentials.
\newblock \emph{Nature Communications}, 6\penalty0 (1), jul 2015.
\newblock \doi{10.1038/ncomms8257}.

\bibitem[Horsley et~al.(2016)Horsley, King, and Philbin]{Horsley_2016}
S~A~R Horsley, C~G King, and T~G Philbin.
\newblock Wave propagation in complex coordinates.
\newblock \emph{Journal of Optics}, 18\penalty0 (4):\penalty0 044016, apr 2016.
\newblock \doi{10.1088/2040-8978/18/4/044016}.

\bibitem[Sounas et~al.(2015)Sounas, Fleury, and Al{\`{u}}]{Sounas_2015}
Dimitrios~L. Sounas, Romain Fleury, and Andrea Al{\`{u}}.
\newblock Unidirectional cloaking based on metasurfaces with balanced loss and
  gain.
\newblock \emph{Physical Review Applied}, 4\penalty0 (1), jul 2015.
\newblock \doi{10.1103/physrevapplied.4.014005}.

\bibitem[Yin and Zhang(2013)]{Yin_2013}
Xiaobo Yin and Xiang Zhang.
\newblock Unidirectional light propagation at exceptional points.
\newblock \emph{Nature Materials}, 12\penalty0 (3):\penalty0 175--177, feb
  2013.
\newblock \doi{10.1038/nmat3576}.

\bibitem[Chong et~al.(2010)Chong, Ge, Cao, and Stone]{Chong_2010}
Y.~D. Chong, Li~Ge, Hui Cao, and A.~D. Stone.
\newblock Coherent perfect absorbers: Time-reversed lasers.
\newblock \emph{Physical Review Letters}, 105\penalty0 (5), jul 2010.
\newblock \doi{10.1103/physrevlett.105.053901}.

\bibitem[Baranov et~al.(2017)Baranov, Krasnok, Shegai, Al{\`{u}}, and
  Chong]{Baranov_2017}
Denis~G. Baranov, Alex Krasnok, Timur Shegai, Andrea Al{\`{u}}, and Yidong
  Chong.
\newblock Coherent perfect absorbers: linear control of light with light.
\newblock \emph{Nature Reviews Materials}, 2\penalty0 (12), oct 2017.
\newblock \doi{10.1038/natrevmats.2017.64}.

\bibitem[King et~al.(2017)King, Horsley, and Philbin]{King_2017}
C.G. King, S.A.R. Horsley, and T.G. Philbin.
\newblock Perfect transmission through disordered media.
\newblock \emph{Physical Review Letters}, 118\penalty0 (16), apr 2017.
\newblock \doi{10.1103/physrevlett.118.163201}.

\bibitem[Hu et~al.(2017)Hu, Wang, Shum, and Chong]{Hu_2017}
Wenchao Hu, Hailong Wang, Perry~Ping Shum, and Y.~D. Chong.
\newblock Exceptional points in a non-hermitian topological pump.
\newblock \emph{Physical Review B}, 95\penalty0 (18), may 2017.
\newblock \doi{10.1103/physrevb.95.184306}.

\bibitem[Feng et~al.(2017)Feng, El-Ganainy, and Ge]{Feng_2017}
Liang Feng, Ramy El-Ganainy, and Li~Ge.
\newblock Non-hermitian photonics based on parity{\textendash}time symmetry.
\newblock \emph{Nature Photonics}, 11\penalty0 (12):\penalty0 752--762, nov
  2017.
\newblock \doi{10.1038/s41566-017-0031-1}.

\bibitem[Shi et~al.(2016)Shi, Dubois, Chen, Cheng, Ramezani, Wang, and
  Zhang]{Shi_2016}
Chengzhi Shi, Marc Dubois, Yun Chen, Lei Cheng, Hamidreza Ramezani, Yuan Wang,
  and Xiang Zhang.
\newblock Accessing the exceptional points of parity-time symmetric acoustics.
\newblock \emph{Nature Communications}, 7\penalty0 (1), mar 2016.
\newblock \doi{10.1038/ncomms11110}.

\bibitem[Rivet et~al.(2018)Rivet, Brandstötter, Makris, Lissek, Rotter, and
  Fleury]{Rivet_2018}
Etienne Rivet, Andre Brandstötter, Konstantinos~G. Makris, Herv{\'{e}} Lissek,
  Stefan Rotter, and Romain Fleury.
\newblock Constant-pressure sound waves in non-hermitian disordered media.
\newblock \emph{Nature Physics}, 14\penalty0 (9):\penalty0 942--947, jul 2018.
\newblock \doi{10.1038/s41567-018-0188-7}.

\bibitem[Cho et~al.(2020)Cho, Wen, Park, and Li]{Cho_2020}
Choonlae Cho, Xinhua Wen, Namkyoo Park, and Jensen Li.
\newblock Digitally virtualized atoms for acoustic metamaterials.
\newblock \emph{Nature Communications}, 11\penalty0 (1), jan 2020.
\newblock \doi{10.1038/s41467-019-14124-y}.

\bibitem[Miri and Al{\`{u}}(2019)]{Miri_2019}
Mohammad-Ali Miri and Andrea Al{\`{u}}.
\newblock Exceptional points in optics and photonics.
\newblock \emph{Science}, 363\penalty0 (6422), jan 2019.
\newblock \doi{10.1126/science.aar7709}.

\bibitem[Wiersig(2016)]{Wiersig_2016}
Jan Wiersig.
\newblock Sensors operating at exceptional points: General theory.
\newblock \emph{Physical Review A}, 93\penalty0 (3), mar 2016.
\newblock \doi{10.1103/physreva.93.033809}.

\bibitem[Hodaei et~al.(2017)Hodaei, Hassan, Wittek, Garcia-Gracia, El-Ganainy,
  Christodoulides, and Khajavikhan]{Hodaei_2017}
Hossein Hodaei, Absar~U. Hassan, Steffen Wittek, Hipolito Garcia-Gracia, Ramy
  El-Ganainy, Demetrios~N. Christodoulides, and Mercedeh Khajavikhan.
\newblock Enhanced sensitivity at higher-order exceptional points.
\newblock \emph{Nature}, 548\penalty0 (7666):\penalty0 187--191, aug 2017.
\newblock \doi{10.1038/nature23280}.

\bibitem[Uzdin et~al.(2011)Uzdin, Mailybaev, and Moiseyev]{Uzdin_2011}
Raam Uzdin, Alexei Mailybaev, and Nimrod Moiseyev.
\newblock On the observability and asymmetry of adiabatic state flips generated
  by exceptional points.
\newblock \emph{Journal of Physics A: Mathematical and Theoretical},
  44\penalty0 (43):\penalty0 435302, oct 2011.
\newblock \doi{10.1088/1751-8113/44/43/435302}.

\bibitem[Berry and Uzdin(2011)]{Berry_2011}
M~V Berry and R~Uzdin.
\newblock Slow non-hermitian cycling: exact solutions and the stokes
  phenomenon.
\newblock \emph{Journal of Physics A: Mathematical and Theoretical},
  44\penalty0 (43):\penalty0 435303, oct 2011.
\newblock \doi{10.1088/1751-8113/44/43/435303}.

\bibitem[Nada et~al.(2017)Nada, Othman, and Capolino]{Nada_2017}
Mohamed~Y. Nada, Mohamed A.~K. Othman, and Filippo Capolino.
\newblock Theory of coupled resonator optical waveguides exhibiting high-order
  exceptional points of degeneracy.
\newblock \emph{Physical Review B}, 96\penalty0 (18), nov 2017.
\newblock \doi{10.1103/physrevb.96.184304}.

\bibitem[Sayyad and Kunst(2022)]{Sayyad_2022}
Sharareh Sayyad and Flore~K. Kunst.
\newblock Realizing exceptional points of any order in the presence of
  symmetry.
\newblock \emph{Physical Review Research}, 4\penalty0 (2), may 2022.
\newblock \doi{10.1103/physrevresearch.4.023130}.

\bibitem[Draine and Flatau(1994)]{Draine_1994}
Bruce~T. Draine and Piotr~J. Flatau.
\newblock Discrete-dipole approximation for scattering calculations.
\newblock \emph{Journal of the Optical Society of America A}, 11\penalty0
  (4):\penalty0 1491, apr 1994.
\newblock \doi{10.1364/josaa.11.001491}.

\bibitem[Yurkin and Hoekstra(2007)]{Yurkin_2007}
M.A. Yurkin and A.G. Hoekstra.
\newblock The discrete dipole approximation: An overview and recent
  developments.
\newblock \emph{Journal of Quantitative Spectroscopy and Radiative Transfer},
  106\penalty0 (1-3):\penalty0 558--589, jul 2007.
\newblock \doi{10.1016/j.jqsrt.2007.01.034}.

\bibitem[Purcell and Pennypacker(1973)]{Purcell_1973}
Edward~M. Purcell and Carlton~R. Pennypacker.
\newblock Scattering and absorption of light by nonspherical dielectric grains.
\newblock \emph{The Astrophysical Journal}, 186:\penalty0 705, dec 1973.
\newblock \doi{10.1086/152538}.

\bibitem[Landy and Smith(2014)]{Landy_2014}
Nathan Landy and David~R. Smith.
\newblock Two-dimensional metamaterial device design in the discrete dipole
  approximation.
\newblock \emph{Journal of Applied Physics}, 116\penalty0 (4):\penalty0 044906,
  jul 2014.
\newblock \doi{10.1063/1.4891295}.

\bibitem[Capers et~al.(2021)Capers, Boyes, Hibbins, and Horsley]{Capers_2021}
James~R. Capers, Stephen~J. Boyes, Alastair~P. Hibbins, and Simon A.~R.
  Horsley.
\newblock Designing the collective non-local responses of metasurfaces.
\newblock \emph{Communications Physics}, 4\penalty0 (1), sep 2021.
\newblock \doi{10.1038/s42005-021-00713-1}.

\bibitem[Baker et~al.(2021)Baker, Liu, and McLeod]{Baker_2021}
Maryam Baker, Weilin Liu, and Euan McLeod.
\newblock Accurate and fast modeling of scattering from random arrays of
  nanoparticles using the discrete dipole approximation and angular spectrum
  method.
\newblock \emph{Optics Express}, 29\penalty0 (14):\penalty0 22761, jul 2021.
\newblock \doi{10.1364/oe.431754}.

\bibitem[Zubko et~al.(2010)Zubko, Petrov, Grynko, Shkuratov, Okamoto, Muinonen,
  Nousiainen, Kimura, Yamamoto, and Videen]{Zubko_2010}
Evgenij Zubko, Dmitry Petrov, Yevgen Grynko, Yuriy Shkuratov, Hajime Okamoto,
  Karri Muinonen, Timo Nousiainen, Hiroshi Kimura, Tetsuo Yamamoto, and Gorden
  Videen.
\newblock Validity criteria of the discrete dipole approximation.
\newblock \emph{Applied Optics}, 49\penalty0 (8):\penalty0 1267, mar 2010.
\newblock \doi{10.1364/ao.49.001267}.

\bibitem[Salary et~al.(2019)Salary, Jafar-Zanjani, and Mosallaei]{Salary_2019}
Mohammad~Mahdi Salary, Samad Jafar-Zanjani, and Hossein Mosallaei.
\newblock Nonreciprocal optical links based on time-modulated nanoantenna
  arrays: Full-duplex communication.
\newblock \emph{Physical Review B}, 99\penalty0 (4), jan 2019.
\newblock \doi{10.1103/physrevb.99.045416}.

\bibitem[DeVoe(1964)]{DeVoe_1964}
Howard DeVoe.
\newblock Optical properties of molecular aggregates. i. classical model of
  electronic absorption and refraction.
\newblock \emph{The Journal of Chemical Physics}, 41\penalty0 (2):\penalty0
  393--400, jul 1964.
\newblock \doi{10.1063/1.1725879}.

\bibitem[Euler(1735)]{Euler_1735}
Leonhard Euler.
\newblock Solutio problematis ad geometriam situs pertinentis.
\newblock \emph{Commentarii academiae scientiarum Petropolitanae}, 8\penalty0
  (24):\penalty0 128--140, aug 1735.

\bibitem[Foulds(1992)]{Foulds_1992}
L.~R. Foulds.
\newblock \emph{Graph Theory Applications}.
\newblock Springer New York, 1992.
\newblock \doi{10.1007/978-1-4612-0933-1}.

\bibitem[Coates(1959)]{Coates_1959}
C.~Coates.
\newblock Flow-graph solutions of linear algebraic equations.
\newblock \emph{{IRE} Transactions on Circuit Theory}, 6\penalty0 (2):\penalty0
  170--187, 1959.
\newblock \doi{10.1109/tct.1959.1086537}.

\bibitem[Brualdi and Cvetkovic(2008)]{Brualdi_2008}
Richard~A. Brualdi and Dragos Cvetkovic.
\newblock \emph{A Combinatorial Approach to Matrix Theory and Its
  Applications}.
\newblock Chapman and Hall/{CRC}, aug 2008.
\newblock \doi{10.1201/9781420082241}.

\bibitem[West(2001)]{West_2001}
Douglas~Brent West.
\newblock \emph{Introduction to graph theory}.
\newblock Pearson, 2001.

\bibitem[Bollob{\'{a}}s(1979)]{Bollob_s_1979}
B{\'{e}}la Bollob{\'{a}}s.
\newblock \emph{Graph Theory}.
\newblock Springer New York, 1979.
\newblock \doi{10.1007/978-1-4612-9967-7}.

\bibitem[Greub(1963)]{Greub_1963}
Werner~H. Greub.
\newblock \emph{Linear Algebra}.
\newblock Springer Berlin Heidelberg, 1963.
\newblock \doi{10.1007/978-3-662-01545-2}.

\bibitem[Greenman(1976)]{Greenman_1976}
J.~V. Greenman.
\newblock Graphs and determinants.
\newblock \emph{The Mathematical Gazette}, 60\penalty0 (414):\penalty0
  241--246, dec 1976.
\newblock \doi{10.2307/3615432}.

\bibitem[Newton(1982)]{Newton_1982}
Roger~G. Newton.
\newblock \emph{Scattering Theory of Waves and Particles}.
\newblock Springer Berlin Heidelberg, 1982.
\newblock \doi{10.1007/978-3-642-88128-2}.

\bibitem[Horn and Johnson(2012)]{Horn_2012}
Roger~A. Horn and Charles~R. Johnson.
\newblock \emph{Matrix Analysis}.
\newblock Cambridge University Press, oct 2012.
\newblock \doi{10.1017/cbo9781139020411}.

\bibitem[Demange and Graefe(2011)]{Demange_2011}
Gilles Demange and Eva-Maria Graefe.
\newblock Signatures of three coalescing eigenfunctions.
\newblock \emph{Journal of Physics A: Mathematical and Theoretical},
  45\penalty0 (2):\penalty0 025303, dec 2011.
\newblock \doi{10.1088/1751-8113/45/2/025303}.

\bibitem[Wiersig(2014)]{Wiersig_2014}
Jan Wiersig.
\newblock Enhancing the sensitivity of frequency and energy splitting detection
  by using exceptional points: Application to microcavity sensors for
  single-particle detection.
\newblock \emph{Physical Review Letters}, 112\penalty0 (20), may 2014.
\newblock \doi{10.1103/physrevlett.112.203901}.

\bibitem[Lin et~al.(2016)Lin, Pick, Lon{\v{c}}ar, and Rodriguez]{Lin_2016}
Zin Lin, Adi Pick, Marko Lon{\v{c}}ar, and Alejandro~W. Rodriguez.
\newblock Enhanced spontaneous emission at third-order dirac exceptional points
  in inverse-designed photonic crystals.
\newblock \emph{Physical Review Letters}, 117\penalty0 (10), aug 2016.
\newblock \doi{10.1103/physrevlett.117.107402}.

\bibitem[Dembowski et~al.(2001)Dembowski, Gräf, Harney, Heine, Heiss, Rehfeld,
  and Richter]{Dembowski_2001}
C.~Dembowski, H.-D. Gräf, H.~L. Harney, A.~Heine, W.~D. Heiss, H.~Rehfeld, and
  A.~Richter.
\newblock Experimental observation of the topological structure of exceptional
  points.
\newblock \emph{Physical Review Letters}, 86\penalty0 (5):\penalty0 787--790,
  jan 2001.
\newblock \doi{10.1103/physrevlett.86.787}.

\bibitem[Chen et~al.(2017)Chen, Özdemir, Zhao, Wiersig, and Yang]{Chen_2017}
Weijian Chen, {\c{S}}ahin~Kaya Özdemir, Guangming Zhao, Jan Wiersig, and Lan
  Yang.
\newblock Exceptional points enhance sensing in an optical microcavity.
\newblock \emph{Nature}, 548\penalty0 (7666):\penalty0 192--196, aug 2017.
\newblock \doi{10.1038/nature23281}.

\bibitem[Banerjee et~al.(2023)Banerjee, Jaiswal, Manjunath, and
  Narayan]{banerjee2023tropical}
Ayan Banerjee, Rimika Jaiswal, Madhusudan Manjunath, and Awadhesh Narayan.
\newblock A tropical geometric approach to exceptional points.
\newblock \emph{Arxiv}, jan 2023.
\newblock \doi{10.48550/arXiv.2301.13485}.

\bibitem[Brand(1964)]{Brand_1964}
Louis Brand.
\newblock The companion matrix and its properties.
\newblock \emph{The American Mathematical Monthly}, 71\penalty0 (6):\penalty0
  629--634, jun 1964.
\newblock \doi{10.1080/00029890.1964.11992294}.

\bibitem[Ding(2010)]{Ding_2010}
Feng Ding.
\newblock Transformations between some special matrices.
\newblock \emph{Computers {\&} Mathematics with Applications}, 59\penalty0
  (8):\penalty0 2676--2695, apr 2010.
\newblock \doi{10.1016/j.camwa.2010.01.036}.

\bibitem[Bellman(1997)]{Bellman_1997}
Richard Bellman.
\newblock \emph{Introduction to Matrix Analysis, Second Edition}.
\newblock Society for Industrial and Applied Mathematics, jan 1997.
\newblock \doi{10.1137/1.9781611971170}.

\bibitem[Scali et~al.(2021)Scali, Anders, and Correa]{Scali_2021}
Stefano Scali, Janet Anders, and Luis~A. Correa.
\newblock Local master equations bypass the secular approximation.
\newblock \emph{Quantum}, 5:\penalty0 451, may 2021.
\newblock \doi{10.22331/q-2021-05-01-451}.

\bibitem[Rotter(2010)]{rotter2010role}
Ingrid Rotter.
\newblock The role of exceptional points in quantum systems.
\newblock \emph{Arxiv}, nov 2010.
\newblock \doi{10.48550/arXiv.1011.0645}.

\bibitem[Tang et~al.(2020)Tang, Jiang, Ding, Xiao, Zhang, Chan, and
  Ma]{Tang_2020}
Weiyuan Tang, Xue Jiang, Kun Ding, Yi-Xin Xiao, Zhao-Qing Zhang, C.~T. Chan,
  and Guancong Ma.
\newblock Exceptional nexus with a hybrid topological invariant.
\newblock \emph{Science}, 370\penalty0 (6520):\penalty0 1077--1080, nov 2020.
\newblock \doi{10.1126/science.abd8872}.

\bibitem[G{\"u}nther et~al.(2007)G{\"u}nther, Rotter, and
  Samsonov]{Gunther_2007}
Uwe G{\"u}nther, Ingrid Rotter, and Boris~F Samsonov.
\newblock Projective hilbert space structures at exceptional points.
\newblock \emph{Journal of Physics A: Mathematical and Theoretical},
  40\penalty0 (30):\penalty0 8815--8833, jul 2007.
\newblock \doi{10.1088/1751-8113/40/30/014}.

\bibitem[Ma and Edelman(1998)]{Ma_1998}
Yanyuan Ma and Alan Edelman.
\newblock Nongeneric eigenvalue perturbations of jordan blocks.
\newblock \emph{Linear Algebra and its Applications}, 273\penalty0
  (1-3):\penalty0 45--63, apr 1998.
\newblock \doi{10.1016/s0024-3795(97)00342-x}.

\bibitem[Markel(2019)]{Markel_2019}
Vadim~A. Markel.
\newblock Extinction, scattering and absorption of electromagnetic waves in the
  coupled-dipole approximation.
\newblock \emph{Journal of Quantitative Spectroscopy and Radiative Transfer},
  236:\penalty0 106611, oct 2019.
\newblock \doi{10.1016/j.jqsrt.2019.106611}.

\bibitem[Sweeney et~al.(2019)Sweeney, Hsu, Rotter, and Stone]{Sweeney_2019}
William~R. Sweeney, Chia~Wei Hsu, Stefan Rotter, and A.~Douglas Stone.
\newblock Perfectly absorbing exceptional points and chiral absorbers.
\newblock \emph{Physical Review Letters}, 122\penalty0 (9), mar 2019.
\newblock \doi{10.1103/physrevlett.122.093901}.

\bibitem[Wang et~al.(2021)Wang, Sweeney, Stone, and Yang]{Wang_2021}
Changqing Wang, William~R. Sweeney, A.~Douglas Stone, and Lan Yang.
\newblock Coherent perfect absorption at an exceptional point.
\newblock \emph{Science}, 373\penalty0 (6560):\penalty0 1261--1265, sep 2021.
\newblock \doi{10.1126/science.abj1028}.

\bibitem[Krasnok et~al.(2019)Krasnok, Baranov, Li, Miri, Monticone, and
  Al{\'{u}}]{Krasnok_2019}
Alex Krasnok, Denis Baranov, Huanan Li, Mohammad-Ali Miri, Francesco Monticone,
  and Andrea Al{\'{u}}.
\newblock Anomalies in light scattering.
\newblock \emph{Advances in Optics and Photonics}, 11\penalty0 (4):\penalty0
  892, dec 2019.
\newblock \doi{10.1364/aop.11.000892}.

\bibitem[Krasnok et~al.(2021)Krasnok, Nefedkin, and Alu]{Krasnok_2021}
Alex Krasnok, Nikita Nefedkin, and Andrea Alu.
\newblock Parity-time symmetry and exceptional points [electromagnetic
  perspectives].
\newblock \emph{{IEEE} Antennas and Propagation Magazine}, 63\penalty0
  (6):\penalty0 110--121, dec 2021.
\newblock \doi{10.1109/map.2021.3115766}.

\bibitem[Ghaemi-Dizicheh and Schomerus(2021)]{Ghaemi_Dizicheh_2021}
Hamed Ghaemi-Dizicheh and Henning Schomerus.
\newblock Compatibility of transport effects in non-hermitian nonreciprocal
  systems.
\newblock \emph{Physical Review A}, 104\penalty0 (2), aug 2021.
\newblock \doi{10.1103/physreva.104.023515}.

\bibitem[Zhang et~al.(2022)Zhang, Zhang, Lu, and Chen]{Zhang_2022}
Xiujuan Zhang, Tian Zhang, Ming-Hui Lu, and Yan-Feng Chen.
\newblock A review on non-hermitian skin effect.
\newblock \emph{Advances in Physics: X}, 7\penalty0 (1), aug 2022.
\newblock \doi{10.1080/23746149.2022.2109431}.

\bibitem[Xin et~al.(2023)Xin, Song, Wu, Lin, Zhu, and Li]{Xin_2023}
Haoran Xin, Wange Song, Shengjie Wu, Zhiyuan Lin, Shining Zhu, and Tao Li.
\newblock Manipulating the non-hermitian skin effect in optical ring
  resonators.
\newblock \emph{Physical Review B}, 107\penalty0 (16), apr 2023.
\newblock \doi{10.1103/physrevb.107.165401}.

\end{thebibliography}

\end{document}